\documentclass[11pt,letterpaper]{JHEP3}
\pdfoutput=1

\usepackage{epsfig,multicol,bbm}
\usepackage{graphicx}
\usepackage{amsfonts}
\usepackage{amsmath,amssymb,epsfig}
\usepackage{mqcommands}
\newcommand{\mo}{\mathcal O}
\newcommand{\iads}{\int_{\mbox{\tiny $AdS$}}}
\newcommand{\ib}{\int_{\partial\mbox{\tiny $AdS$}}}
\newcommand{\poc}[2]{\left(#1\right)_{#2}}

\newcommand{\reef}[1]{(\ref{#1})}
\renewcommand{\eqref}[1]{(\ref{#1})}

\newcommand{\gt}[1]{\Gamma\left(\frac{ #1}{2}\right)}
\newcommand{\gn}[1]{\Gamma\left( #1\right)}

\newcommand{\X}[2]{X_{\mbox{\tiny #1#2}}}
\newcommand{\dd}[1]{\Delta_{#1}}
\newcommand{\bs}[1]{\bar s_{#1}}
\newcommand{\de}[2]{\delta_{#1 #2}}
\newcommand{\bst}[1]{\bar s_{#1}^2}

\title{Towards Feynman rules for Mellin amplitudes in AdS/CFT}

\author{Miguel F. Paulos$^{a}$\\
$^a$ {\it Laboratoire de Physique Th\'eorique et Hautes Energies, CNRS UMR 7589,\\ Universit\'e Pierre et Marie Curie, 4 place Jussieu, 75252 Paris Cedex 05, France}\\

\vskip .5cm

{\rm E-mail:}\ \ {\
mpaulos@lpthe.jussieu.fr}}

\abstract{
We investigate the use of the embedding formalism and the Mellin transform in the calculation of tree-level conformal correlation functions in $AdS$/CFT. 
We evaluate 5- and 6-point Mellin amplitudes in $\phi^3$ theory and even a 12-pt diagram in $\phi^4$ theory, enabling us to conjecture a set of Feynman rules for scalar Mellin amplitudes. The general vertices are given in terms of Lauricella generalized hypergeometric functions. We also show how to use the same combination of Mellin transform and embedding formalism for amplitudes involving fields with spin. The complicated tensor structures which usually arise can be written as certain operators acting as projectors on much simpler index structures - essentially the same ones appearing in a flat space amplitude. Using these methods we are able to evaluate a four-point current diagram with current exchange in Yang-Mills theory.
}

\preprint{}

\begin{document}{\vskip 1cm}

\section{Introduction}

Witten diagrams \cite{Witten:1998qj} provide us with the means for calculating correlation functions \cite{Gubser:1998bc} of strongly coupled conformal field theories with a gravity dual \cite{Maldacena:1997re,Aharony:1999ti}. However, in spite of significant progress \cite{Freedman:1998tz,Freedman:1998bj,D'Hoker:1999jc,D'Hoker:1999ni}, such calculations are in general quite cumbersome to perform. As it stands, the state of the art is the computation of four point functions involving different kinds of exchanged fields in type IIB supergravity\footnote{See also the works \cite{Bartels:2009jb, Bartels:2009sc,Hatta:2010kt} 
where correlators of currents are calculated in certain limits.}
\cite{D'Hoker:1999pj,Uruchurtu:2007kq,Uruchurtu:2008kp,Uruchurtu:2011wh}, and a stress-tensor three point function \cite{Arutyunov:1999nw}. The latter constitutes an especially heroic effort, due to the complicated tensor structures required for conformal invariance of the three-point function \cite{Schreier:1971um,Osborn:1993cr}.

Such calculations are usually performed in coordinate space. An obvious question is whether changing basis could lead to simplifications. The first guess is momentum space, but this doesn't lead to any major simplifications - perhaps the reason is simply that such a transformation does not take into account the symmetries of $AdS$ space, but only of its boundary. As it turns out that a more appropriate basis does exist: instead of the Fourier transform one should really be working with the Mellin transform \cite{Mack:2009mi,Mack:2009gy,Penedones:2010ue}

The Mellin transform is very natural from a conformal field theory perspective. To see this consider the four-point function of a scalar fields $\mathcal O_i$ of conformal dimension $\Delta_i$. By using the OPE in the $12$ channel say, we can write
	\bea
	&& \left \langle \mathcal O_{\dd 1}(x_1) \mathcal O_{\dd 1}(x_2) \mathcal O_{\dd 1}(x_3) \mathcal O_{\dd 1}(x_4)\right \rangle =
	 \int \frac{\ud c}{2\pi i} g^{(12)(34)}(c) \nonumber \\
	&& \int \ud^d x \left\langle \mo(x_1) \mo(x_2) \phi_{h+c}(x)\right \rangle
	\left\langle  \phi_{h-c}(x) \mo(x_3) \mo(x_4)\right \rangle+\ldots 		\label{cft4pt}
	\eea
where the $\ldots$ represent contributions of fields with spin appearing in the OPE, $\phi_{h\pm c}$ is a scalar field of unphysical dimension $h\pm c$, and $g^{(12)(34)}(c)$ contains the information about which scalar fields appear in the OPE, through its pole structure. The three point functions appearing above are uniquely fixed by conformal symmetry, say
	\be
	\left\langle \mo_{\Delta_1}(x_1) \mo_{\Delta_2}(x_2) \mo_{\Delta_3}(x)\right \rangle=
	C_{\Delta_1, \Delta_2, \Delta_3} \prod_{i<j}^3 (x_i-x_j)^{-2\tilde \Delta_{ij}}
	\ee
with {\em e.g.} $\tilde \Delta_{12}=\frac 12 (\Delta_1+\Delta_2-\Delta_3)$ and $C_{\Delta_1, \Delta_2, \Delta_3}$ is a constant which contains information about the dynamics. Therefore the integral becomes
	\bea
	&& \left \langle \mathcal O(x_1) \mathcal O(x_2) \mathcal O(x_3) \mathcal O(x_4)\right \rangle =
	 \int \frac{\ud c}{2\pi i} g^{(12)(34)}(c) C_{\Delta_1, \Delta_2, \, h+c} C_{\Delta_3, \Delta_4, \, h-c}\nonumber \\
	&& (x_1-x_2)^{-\left(\Delta_1+\Delta_2-(h+c)\right)}
	(x_3-x_4)^{-\left(\Delta_3+\Delta_4-(h+c)\right)}
	\int \ud^d x \prod_{i=1}^4 (x-x_i)^{-\delta_i}
+\ldots \label{intcft4pt}	
	\eea
with
	\bea
	\delta_1=\frac 12 (\Delta_1+h+c-\Delta_2), \quad \delta_2=\frac 12 (\Delta_2+h+c-\Delta_1), \nonumber \\
	\delta_3=\frac 12 (\Delta_3+h-c-\Delta_4), \quad \delta_2=\frac 12 (\Delta_4+h-c-\Delta_3),
	\eea

To perform the $x$ integral in \reef{intcft4pt} the standard procedure is to introduce Schwinger parameters to exponentiate the denominators. The $x$ integration becomes trivial, and the Schwinger integrations can be performed via Symanzik's star formula \cite{Symanzik:1972wj}, as we discuss in appendix \ref{Symanzik}. The net result is that
	\bea
	\pi^{-d/2}\int \ud^d x \prod_{i=1}^4 (x-x_i)^{-\delta_i} \Gamma(\delta_i)=
	\int \ud \delta_{ij} \prod_{i<j}^4 \Gamma(\tilde \delta_{ij}) (x_i-x_j)^{-2\tilde \delta_{ij}}
	\eea
where the $n(n-3)/2$ independent parameters $\tilde \delta_{ij}$ satisfy the constraints $\sum_{i\neq j} \tilde \delta_{ij}=\delta_j$. In this way, we have passed from integrations in coordinate space to integrations in the Mellin space. 

Generically, any conformal field theory correlation function of scalars with dimensions $\Delta_i$ can be written in the Mellin representation as \cite{Mack:2009mi}:
	\be
		A(x_1,x_2,\ldots, x_n)=\frac{\mathcal N}{(2\pi i)^{\frac 12 n(n-3)}}
		\int \ud \delta_{ij}\, \, M(\delta_{ij}) \, 		  \prod_{i<j}^n\Gamma(\delta_{ij})(x_i-x_j)^{-2\delta_{ij}}. \label{Mellin}
	\ee 
The normalization constant $\mathcal N$ will be fixed later. The object $M(\delta_{ij})$ is the Mellin amplitude, which depends on a set of $n(n-3)/2$ parameters $\delta_{ij}$ equal in number to the number of independent cross-ratios \footnote{As long as this number is smaller than $n\times d$, the number of maximally independent components of $n$-dimensional vectors.}

. These parameters satisfy the constraints
	\be
		\sum_j \de i j=\dd i, \qquad \delta_{ii}=0
	\ee
which may be solved by introducing a set of $d$-dimensional vectors $k_i$ satisfying
\be
-k_i^2=\Delta_i, \qquad \sum_i k_i=0
\ee
in terms of which $\delta_{ij}=k_i\cdot k_j$. It is also useful to introduce the ``Mandelstam invariants''
	\be
		s_{i_1 i_2 \ldots i_p}=-\left(\sum_{m=1}^p k_{i_m}\right)^2
		=\sum_{m=1}^p \Delta_{i_m}-2 \sum_{i_k<i_l} \delta_{i_k j_l}, \label{mandel}
	\ee
which imply for instance $s_{ij}=-(k_i+k_j)^2=\Delta_i+\Delta_j-2 \delta_{ij}$.

Mellin amplitudes have very simple analytic properties. The scalar four-point function for instance, has an infinite set of simple poles in the $s$-channel at $s_{12}=\Delta_k-s_k+2n$, where $\Delta_k$, $s_k$ are the conformal dimension and spin respectively of a field appearing in the OPE, and $n$ is a positive integer. The residues of the satellite poles, that is those with $n\neq 0$, are completely fixed by conformal symmetry in terms of the leading $n=0$ pole. Further, validity of the OPE requires factorisation: the residue of the leading pole splits into the product of two factors, one pertaining only to fields $12$ and the other to fields $34$. 

In the paper \cite{Penedones:2010ue}, the Mellin formalism was used to study CFT correlation functions computed in the $AdS$/CFT context, with promising results. For instance, contact interactions have simply polynomials as their Mellin amplitudes, in contrast to the complicated $D$-functions which appear in coordinate space. Even the dreaded stress-tensor exchange diagram reduces to a simple rational function for the case of minimally coupled massless scalars.
The simple analytic properties of Mellin amplitudes also make clear which operators are propagating throughout a given Witten diagram: double-trace operators corresponding to the fusion of external legs are captured by the explicit gamma functions in the Mellin representation, whereas single-trace operators and their descendants corresponding to internal lines or bulk-to-bulk propagators, appear as simple poles of the Mellin amplitude,.

In this paper we continue to investigate the properties of $AdS$/CFT correlation functions in the Mellin representation. We shall do this on two fronts. Firstly by evaluating higher point amplitudes in purely scalar theory, that is, where no other fields other than scalars propagate in a Witten diagram. Secondly by computing correlation functions of operators with spin such as currents and stress-tensors. In both cases it will be invaluable to use the embedding formalism \cite{Dirac:1936fq, Weinberg:2010fx,Penedones:2007ns}. The main idea is 
to think of $AdS_{d+1}$ space as embedded in flat Minkowski space $M_{d+2}$, with metric $\eta^{MN}$. $AdS$ coordinate vectors $X^M$ satisfy $X\cdot X=-R^2$ whereas $AdS$ boundary coordinates $P^M$ are defined by $P^2=0$, $P\simeq \alpha P, \alpha>0$. With the two-pronged approach of using embedding formalism and Mellin transforms, the computation of correlation functions simplifies dramatically. 

\subsection{Summary of results}

An intriguing possibility raised by the work of \cite{Penedones:2010ue} is the existence of Feynman rules for Mellin amplitudes. Indeed, the Mellin amplitude for a scalar four point function in $\phi^3$ theory takes the simple form
	\be
	M_4\simeq \sum_{n=0}^{+\infty}\left( \frac{V_{\Delta,n}^2}{s_{12}-\Delta-2n}+\frac{V_{\Delta,n}^2}{s_{13}-\Delta-2n}+\frac{V_{\Delta,n}^2}{s_{14}-\Delta-2n}\right)
	\ee
The vertex $V_{\Delta,n}$ essentially describes the three point function of two scalars and a descendant field at level $n$. The above is remarkably similar to a flat space scattering amplitude, and indeed it becomes one for high enough energies as compared to the dimensions $\Delta$. In this work we shall present strong evidence that at least for scalar theory, it is possible to write down a set of Feynman rules for Mellin amplitudes. More precisely, we compute 5-pt, 6-pt and even a 12-pt diagram in scalar theory and check that the rules hold. These calculations also allow us to read off the vertices $V$ when more than one descendant fields are involved. In $\phi^3$ theory we need at least three internal lines (bulk-to-bulk propagators) to see three descendant fields interacting, and in $\phi^4$ theory we need four such lines. 
Our computations are consistent with the existence of a set of Feynman rules for Mellin diagrams, which are given in the following.

\vspace{0.5 cm}

{\bf Conjecture (Feynman rules for Mellin amplitudes):} 
\begin{itshape}
Consider a tree-level Witten diagram involving only scalar fields, consisting of a set of external (bulk to boundary) and internal (bulk to bulk) lines, and vertices connecting them. The corresponding Mellin amplitude is constructed as follows:
	\begin{itemize}
	\item To every line associate momentum $k_j$. Momentum of external lines satisfy $-k_i^2=\Delta_i$. Momentum conservation must hold for the whole amplitude, and at every vertex. 
	\item To every internal line corresponding to a scalar of conformal dimension $\delta_k$, assign an integer $n_k$ and a propagator:
	\be
	\frac{1}{2 n_j! \Gamma(1+\delta_j+n_j-h)}\,
	\frac{-1}{+k_j^2+(\delta_j+2 n_j)}
	\ee
	\item In $g^{(m)} \phi^m$ theory, the vertex connecting lines with dimension $\Delta_i$, integers $n_i$, is given by
	\bea
 && V^{\dd 1\ldots \dd m}_{[n_1,\ldots, n_m]}=g^{(m)}\, \gt{\sum_i \dd i-2h}
	\left(\prod_{i=1}^n 	\left(1-h+\dd i\right)_{n_i}\right)\nonumber \\
 && F_A^{(m)}\left(\frac {\sum_{i=1}^n \dd i-\!2h}2,\left\{-\!n_1,\ldots,-\!n_m\right\},\left\{1\!+\!\dd1\!-\!h,\ldots,1\!+\!\dd m\!-\!h\right\};1,\ldots,1\right) \hspace{1 cm} \ \ \  
	\eea
	where $(a)_m$ is the Pochhammer symbol and $F_A^{(m)}$ is the Lauricella function of $m$ variables\footnote{The definition is given in equation \reef{lauricella}. Also, see references \cite{Lauricella, Srivastava, mathworld}.}
	\item The Mellin amplitude is obtained by summing over all non-zero integer $n_i$. 
	\end{itemize}
\end{itshape} 

If this conjecture is correct, then correlation functions in the purely scalar sector are completely solved at tree level (other kinds of interactions, such as those including derivatives, can be easily included \cite{Penedones:2010ue}). A proof of these rules will require a better understanding of how lower-point Mellin amplitudes are combined into higher point ones.

An important result in this work, is a simplified formalism for the calculation of correlation functions of objects with indices, such as currents and stress-tensors. We shall find that the bulk to boundary propagators of these objects can be written as certain differential operators $D^{MA}$ acting on scalar propagators. For instance the three-point current Mellin amplitude $M_3^{M_1 M_2 M_3}$ may be written schematically as
	\be
	M_3^{M_1 M_2 M_3}=D^{M_1 A_1}D^{M_2 A_2}D^{M_3 A_3} \tilde M_{A_1 A_2 A_3}.
	\ee
The $D$ operators act as projectors, taking the reduced Mellin amplitude $\tilde M$ onto a conformally invariant subspace. As such, the reduced Mellin amplitude $\tilde M$ is dramatically simpler then the full amplitude. In particular its tensorial structure is essentially the same one that would appear in a flat space scattering amplitude, upon certain identifications.
This simplification holds for arbitrary $n$-point functions, of fields with arbitrary spin. In particular, in this paper we shall carry out as an example the calculation of a four-current Witten diagram involving current exchange in Yang-Mills theory. With some more work, the four-point function of the stress-tensor should be obtainable, since the difficulties involved are essentially the same that are involved in a flat space scattering calculation.

The usage of the embedding formalism also clarifies the requirements of conformal invariance. Consider for instance the current three-point amplitude,
	\be
	\left \langle J^{M_1}(P_1)J^{M_2}(P_2)J^{M_3}(P_3)\right \rangle
	\ee
where all $P_i^{M_i}$ are $d+2$ dimensional vectors which square to zero. To get the $d$-dimensional amplitude we must pull back the $M_i$ indices to $\mu$ indices in $d$ dimensions. This only makes sense if the $M_i$ indices are transverse \cite{Weinberg:2010fx} , that is, if:
	\be
	P_{i,M_i}\left \langle J^{M_1}(P_1)J^{M_2}(P_2)J^{M_3}(P_3)\right \rangle=0
	\ee
for any $i$. This requirement strongly constrains the form of the amplitude. There are essentially two building blocks
	\bea
	X_{ij}^{M_k}&=& \left(\frac{P_i^{M_k}}{P_i\cdot P_k}-\frac{P_j^{M_k}}{P_i\cdot P_k}\right) \label{xdef}\\
	I^{M_i M_j}&=& \eta^{M_i M_j}-\frac{ P_i^{M_j} P_j^{M_i}}{P_i\cdot P_j}, \label{idef}
	\eea
which satisfy $P_{M_k} X_{ij}^{M_k}=P_{M_i} I^{M_i M_j}=P_{M_j} I^{M_i M_j}=0$.
From these we can construct the tensorial structure of {\em any} conformally invariant amplitude. In our example, we find that the amplitude must take the form
	\bea
	\left \langle J^{M_1}(P_1)J^{M_2}(P_2)J^{M_3}(P_3)\right \rangle
	\propto a X_{12}^{M_3} X_{13}^{M_2} X_{12}^{M_1}+b \left(X_{12}^{M_3} \frac{I^{M_1 M_2}}{P_1 \cdot P_2}+\mbox{perms}\right)
	\eea
which is correct \cite{Osborn:1993cr}. However, the reasoning is more general, and it applies to any $n$-point amplitude of any integer spin field.

The layout of this paper is as follows. In the next section we set up our formalism, describing in detail the embedding formalism, and the form of the bulk-to-bulk and bulk-to-boundary propagators that will be used throughout the paper. In section 3 we review some of the results of \cite{Penedones:2010ue}, computing the Mellin amplitude corresponding to a scalar four-point function in $\phi^3$ theory. This will serve as the starting point and motivation for computing higher point amplitudes, in the quest to understand whether Mellin amplitudes can be described by a set of Feynman rules. In sections 4 and 5 we compute five and six-point amplitudes respectively. The form of the amplitudes is consistent with the Feynman rules we described previously, and we read off the general cubic vertex involving three descendant fields, given in terms of the Lauricella function of three arguments. In section 7 we turn our attention to correlators of spin-1 fields. We start by reproducing in a much simpler fashion several computations which have appeared previously in the literature: namely correlators $\langle J\mo \mo\rangle, \langle JJJ\rangle$ and a current exchange diagram in scalar theory. Putting all the ingredients together we are able to explicitly compute a current 4-point function. 
We finish with a brief discussion of our results and prospects for future work.


\vskip 0.5 cm

{\bf Note:}
While this work was being completed, we became aware of the work of
\cite{PenedonesEtAl}  which partially overlap with some of our results. We thank the authors for granting us access to an early version of their manuscript.

\section{Preliminaries}

\subsection{Embedding formalism}

Throughout this paper we shall make strong use of the embedding formalism. In this formalism, $AdS_{d+1}$ space is seen as a curved surface embedded in flat Minkowski space $M_{d+2}$. The Minkowski space metric is denoted $\eta^{MN}$, and it is written as
	\be
	ds^2=-\ud X^+ \ud X^-+\delta_{m n} \ud X^m \ud X^n.
	\ee
That is, we describe the first two directions with lightcone coordinates. $AdS$ coordinate vectors $X^M$ satisfy $X\cdot X=-R^2$ whereas $AdS$ boundary coordinates $P^M$ are defined by $P^2=0$. We are also free to perform rescalings $P\to \alpha P, \alpha>0$, and as such amplitudes $M(P_i)$ satisfying conformal invariance should also scale: $M(P_i)\to \alpha^\Delta M$ for some $\Delta$.
To fix notation we choose
	\begin{itemize}
		\item $P_i$ - fixed boundary points.
		\item $Q_i$ - boundary points integrated over.
		\item $X_i$ - $AdS$ bulk coordinate.
	\end{itemize}
We will also set throughout the rest of this paper the $AdS$ radius to one. Dependence on this quantity can be recovered by dimensional analysis. Useful parameterizations of $AdS$ and its boundary are
	\be
	X^A(x^a)=\frac{1}{x_0}(1,x_0^2+x^2,x^\mu), \qquad 					P^M(x^\mu)=(1,y^2,y^\mu). \label{param}
	\ee
where $x^\mu$ is a $d$-dimensional vector and $x^2=x^\mu x_\mu$.
In this way we have for instance:
	\bea
	P_{ij}&\equiv& -2 P_i\cdot P_j=(y_i-y_j)^2 \\
	-2 P\cdot X&=& \frac{1}{x_0}(x_0^2+(x-y)^2). \label{px}
	\eea
Objects with indices $T_{A_1 \ldots}$ are tensors in $AdS$ if they satisfy $X^{A_1} T_{A_1 \ldots}=0$ \cite{Cornalba:2009ax,Weinberg:2010fx}. To implement this transversality condition one may use the projector
	\be
	U^{AB}=\eta^{AB}+X^A X^B.
	\ee
It is also useful to know how to write such $d+2$ tensors in terms of $d$-dimensional ones. In other words, we need to be able to pull-back $M$ indices to $\mu$ indices, and this is achieved by use of the objects
	\be
	\zeta_\mu^M(P)=\frac{\partial P^M(y^\mu)}{\partial y^\mu}, \qquad
	\varphi_a^M(X)=\frac{\partial X^M(x^\mu)}{\partial x^a}.
	\ee
Because of the constraints $X^2=-1, P^2=0$, we necessarily have $\zeta_\mu(P) \cdot P=\varphi_a(X)\cdot X=0$. Using the parameterization of $AdS$ and its boundary given in \reef{param}, we find the following useful identities:
	\begin{subequations}\label{ident}
	\begin{align}
	\zeta_\mu(y)\cdot P(y')&= y'_\mu-y_\mu,  \\
	\zeta_\mu(y)\cdot X(x)&=\frac{1}{x_0}(x_\mu-y_\mu), \\
	\varphi_{0}(x)\cdot P(y)&=\frac 12 \frac{(y-x)^2-x_0^2}{x_0^2}  		\\
	\varphi_{\mu}(x)\cdot P(y)&=\frac{1}{x_0}(y_\mu-x_\mu)  \\
	\zeta_\mu(x)\cdot \zeta_\nu(y)&= \eta_{\mu \nu} \\
	\varphi_a(x)\cdot \varphi_b(x)&= g_{ab}.  \\
	\varphi_\mu(x)\cdot \zeta_\nu(y)&=\frac{1}{x_0}\, \eta_{\mu\nu}
	\end{align}
	\end{subequations}
In the following we will label indices that are to be contracted with $\zeta_\mu^M$ as $M,N,P,\ldots$, whereas ``$AdS$'' indices will be labelled $A,B,C\ldots$. This provides a practical distinction between boundary and bulk indices, although in the embedding formalism no such distinction exists.

\subsection{Boundary-bulk propagators}

In the $AdS$/CFT correspondence, conformal correlation functions can be calculated via Witten diagrams \cite{Witten:1998qj}. A typical diagram is shown in figure \ref{scexc}.
\FIGURE[ht]{
	\parbox{10 cm}{
	\centering
	\includegraphics[width=8cm]{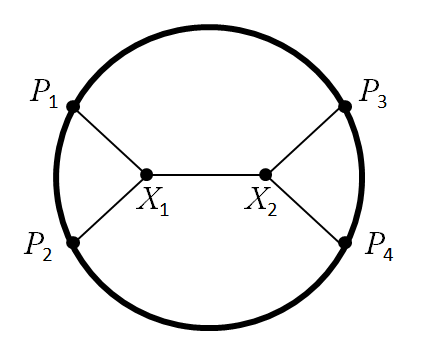}
	\caption{A Witten diagram involving scalar fields.}
		}
	\label{scexc}
}

Such a diagram is made up of three ingredients, namely external lines which connect to the boundary of $AdS$, internal lines, and vertices. The vertices are simple to write down and are easy to read off from the gravitational lagrangian. External lines are bulk-to-boundary propagators, propagating some field perturbation inserted on the boundary into the bulk, and internal lines are bulk-to-bulk propagators. To compute the amplitude we write down a propagator for each line, and integrate over all possible positions of the interaction vertices. In the following we shall give expressions for these propagators in the embedding formalism. Consider first the case where the perturbation corresponds to a scalar operator of conformal dimension $\Delta_i$. Then the propagator can be written as
	\be
		E_i(P,X)=\frac{C_i}{(-2 P\cdot X)^{\Delta_i}}=
		\frac{1}{2\pi^h \Gamma(1+\Delta_i-h)}
		\int_0^{+\infty} \frac{\ud t_i}{t_i} t_i^{\Delta_i}\, 
		e^{2 t_i P\cdot X}. \label{propscalar}
	\ee
Here $i$ is shorthand notation for denoting the field in question and its conformal dimension, and the constants are
	\be
		C_i=\frac{\Gamma(\Delta_i)}{2\pi^h \Gamma(1+\Delta_i-h)}, 
		\qquad h\equiv d/2.
	\ee
It is easy to check using our expression \reef{px} that this reduces to the usual bulk-to-boundary propagator
	\be
	E_i(P,X)\simeq \left( \frac{x_0}{x_0^2+(y-x)^2}\right)^{\Delta_i}.
	\vspace{0.2 cm}
	\ee
However, the most convenient expression to use is the Schwinger parameterized form appearing on the right of \reef{propscalar}, and this will be the one we will be using throughout this paper.

Now consider the bulk-to-boundary propagator of a spin-$1$ field. Such a propagator takes the form\footnote{Our normalization differs from that of \cite{D'Hoker:1999jc} by a factor of $d-1$.}
:
	\be
	E_i^{MA}(P,X)=\frac{1}{2\pi^h \Gamma(1+\Delta_i-h)}
		\int_0^{+\infty} \frac{\ud t_i}{t_i} t_i^{\Delta_i}\, 
		J^{MA}\,e^{2 t_i P\cdot X}.
	\ee
That is, it is given by the product of some tensor structure, to propagate indices, and the scalar propagator of a field of dimension $\Delta_i$. For a Yang-Mills field we will have $\Delta_i=d-1$, but we shall keep it arbitrary for now. Requiring transversality of the tensor structure both in $AdS$ and at its boundary fixes $J^{MA}$:
	\be
	P_M J^{MA}=J^{MA} X_A=0 \quad \Rightarrow \quad J^{MA}=
	\eta^{MA}-\frac{P^A X^M}{P\cdot X}
	\ee
The tensor $J^{MA}$ is a projector, as may be easily checked. It serves the two-fold purpose of making transverse in $X$ objects which contract it on the right, and transverse in $P$ objects which contract it on the left. The reader may check that the propagator written above reduces to the right one for a spin-one field upon the use of the identities \reef{ident}. In fact, using $J^{MA}$, we can write down the bulk-to-boundary propagator for a field of any spin - we just multiply several $J^{MA}$ together and symmetrize appropriately its indices to get the right representation. In particular we can do this to obtain the bulk-to-boundary propagator of the graviton. Before we do this however, we notice that there is an alternative representation of the propagator which will be very useful by using the identity:

	\bea
	\int \frac{\ud t}t\, t^{\Delta} \frac{P^A X^M}{P \cdot X}\, e^{2 t P\cdot X}
	&=&\int \frac{\ud t}t\, t^{\Delta} \frac{(-2\, t)}{\Delta}\, P^A X^M\, e^{2 t P\cdot X}\\
	&=&-\int \frac{\ud t}t\, t^{\Delta} \, \frac{P^A}{\Delta} \frac{\partial}{\partial P^M}\, e^{2 t P\cdot X},
	\eea
we can write $E_i^{MA}(P,X)=D_\Delta^{MA} E_i(P,X)$ with the operator

	\be
	D^{MA}_\Delta\equiv \eta^{MA}+\frac{1}{\Delta} P^A \frac{\partial}{\partial P^M}\equiv \eta^{MA}+\frac{1}{\Delta} P^A \partial_{M_1}. \label{Dop}\vspace{0.3 cm}
	\ee
Similarly, for the spin-2 case, we can also write the bulk-to-boundary propagator in terms of an operator acting on the scalar propagator: 
	\bea
	&&E^{M_1 M_2 A B}_i(P,X)=D_{2,\Delta}^{M_1 M_2 A B} E_{i}(P,X), \nonumber \\
	&&D_{2,\Delta}^{M_1 M_2 A_1 A_2}=
	\eta^{M_1 A_1}\eta^{M_2 A_2}+\frac{1}{\Delta}\left(\eta^{M_1 A_1} P^{A_2}\partial_{M_2}+1\leftrightarrow 2\right)+\frac{P^{A_1}P^{A_2}}{\Delta(\Delta+1)}\partial_{M_1}\partial_{M_2}
	\hspace{0.5 cm}
	\eea
Once again, in applications we should take $\Delta=d$ in the above. 

\subsection{Bulk-to-bulk propagators}
\label{bb props}

Next we consider the bulk-to-bulk propagators. These are associated with internal lines in Witten diagrams. For ease of notation, we will henceforth denote the conformal dimension of fields propagating in these internal lines by a lower case $\delta$, and dimensions of fields on external lines by a capital $\Delta$. Then, for a scalar field of dimension $\delta$, the bulk-to-bulk propagator $G_{BB}(X,Y)$ can be written in the embedding formalism as

	\bea
		G_{BB}(X_1,X_2) &=& 
		\int_{-i \infty}^{+\infty}\frac{\ud c}{2\pi i} f_{\delta,0}(c)				\ib \ud Q\int \widetilde{d^2 s_c}\, 
		e^{2 s Q\cdot X+2 \bar s Q\cdot Y}  
	\eea 
with
	\be
	f_{\delta,0}(c)\equiv \frac{1}{2\pi^{2h}[(\delta-h)^2-c^2]}	\frac{1}{\Gamma(c)\Gamma(-c)}, \qquad 
	\widetilde{\ud^2 s_c}\equiv \frac{\ud s}{s}\frac{\ud \bar s}{\bar s} s^{h+c} \bar s^{h-c}
	\ee
It is remarkable that this can be seen as the product of two boundary to bulk propagators of states with unphysical conformal dimensions $h\pm c$, glued together by the integration over the boundary point $Q$ and over $c$. Bulk to bulk propagators of fields with spin will have the same structure as we shall see shortly. The fact that the dependence of the propagator on $X$ and $Y$ factorises simplifies calculations a great deal, since then an $n$-point amplitude can be obtained by appropriately gluing lower-point amplitudes. In particular, this allows one to ultimately reduce an $n$-point amplitudes to a gluing of three point amplitudes, analogously to (but not quite) BCFW \cite{Britto:2005fq} recursion relations.

The bulk-to-bulk propagator for a spin-one field is written in a similar fashion to the spin-zero case \cite{Balitsky:2011tw}:
	\bea
	G_{BB}^{AB}(X_1,X_2)=\hspace{12 cm} &&\nonumber \\
	\int_{-i \infty}^{+\infty}\frac{\ud c}{2\pi i} f_{\delta,1}(c)
	\ib \ud Q 
	\int \frac{\ud s}s \,s^{h+c} \left(D_{h+c}^{MA}\, e^{2 s Q\cdot X_1}\right)
	\eta_{MN}
	\int \frac{\ud \bar s}{\bar s} \bar s^{h-c} \left(D_{h-c}^{NB}\, e^{2 \bar s Q\cdot X_2}\right) \hspace{1 cm} && \\ \nonumber
	\eea
with 
	\be
	f_{\delta,1}=f_{\delta,0}\, \frac{h^2-c^2}{(\delta-h)^2-c^2}, \qquad 	\delta=d-1 \label{f1}
	\ee
and $D^{MA}_{\Delta}$ the operator defined previously in \reef{Dop}.
Finally, the bulk-to-bulk graviton propagator can be obtained by the replacements \cite{Balitsky:2011tw}
	\bea
	f_{\delta,1}\to f_{\delta,2}&=& f_{\delta,0}\, [(h+1)^2-c^2]\\
	D^{MA}&\to & D^{M_1 M_2 A_1 A_2} \\
	\eta_{MN}&\to & \mathcal E_{M_1 M_2, N_1 N_2}
	\eea
with $\mathcal E$ given by
	\be
	\mathcal E_{M_1 M_2 N_1 N_2} \equiv \, \frac 12\left(\eta_{M_1 N_1}\eta_{M_2 N_2}
	+\eta_{M_1 N_2}\eta_{M_2 N_1}\right)
	-\frac{1}{d} \eta_{M_1 M_2}\eta_{N_1 N_2}.
	\ee
The appearance of $d$ instead of $d+2$ in the above will be explained in section \reef{confinv}. For now it is sufficient to notice that in order to get the correct $d$ dimensional index structure we must have $\mathcal E$ of this form.

%
%
\section{Warm-up: 3 and 4-point scalar correlation functions}%
%
%
\subsection{3-point vertex}
Now that we have expressions for all the propagators, we are ready to compute some amplitudes. We will see that using both the embedding formalism and the Schwinger parameterized form of the propagators naturally leads to the appearance of the Mellin transform of the amplitudes, as well as simplifying considerably the calculations.


\FIGURE[ht]{
	\parbox{10 cm}{
	\centering
	\includegraphics[width=7cm]{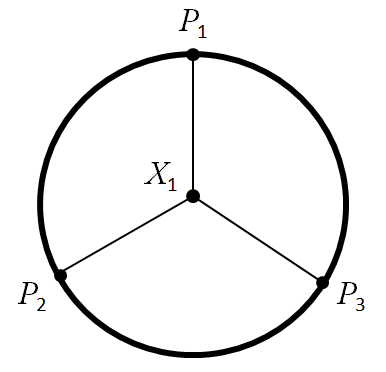}
	\caption{Scalar three-point function.}
	}
	\label{scalar3pt}
}



As a warm-up, consider first a simple theory of massive scalars in $AdS_{d+1}$ interacting via a cubic potential:
	\be
		S_\phi=\int \ud^{d+1} x \sqrt{g}
		\left(\sum_i \frac 12 (\partial \phi_i)^2+
		\frac 12 m_i^2 \phi_i^2
		+\frac{g}{3!}\, \left(\sum_i\phi_i\right)^3 \right).
	\ee
The conformal dimension of the operator $\mathcal O_i$ dual to $\phi_i$ is then $\Delta_i=h\pm \sqrt{h^2+m_i^2}$. 
We start by calculating a scalar three point function, described by the Witten diagram of figure \ref{scalar3pt}.
To each leg connected to the boundary we associate a boundary to bulk propagator $E_i$. We are then instructed to integrate over the interaction point in the bulk of $AdS$, so that the overall amplitude is given by
	\bea 
	A(1,2,3)\equiv \langle \mathcal O_1(P_1) \mathcal O_2(P_2) \mathcal O_3(P_3)\rangle =
	g\iads \ud X \, E_1(P_1,X) E_2(P_2,X)E_3(P_3,X),&& \nonumber \\
	=g\mathcal E_3 \int_0^{+\infty}
	\prod_{i=1}^3 \frac{\ud t_i}{t_i}\, t_i ^{\Delta_i}
	\iads \ud X \exp\left({2(t_1 P_1+t_2 P_2+t_3 P_3)\cdot X}\right)&& 
	\eea
with $\mathcal E_3=\prod_{i=1}^3 \frac{C_i}{\Gamma(\Delta_i)}$.
To proceed we use the result \reef{Xint}, whereupon we obtain
	\be
	A(1,2,3)= g\, \pi^h \, \mathcal E_3\,\gt{\sum_i^n \Delta_i-2h}  \int \prod_{i=1}^3 \frac{\ud t_i}{t_i}\, t_i ^{\Delta_i}
	\exp\left(-t_1 t_2 P_{12}-t_1 t_3 P_{13}-t_2 t_3 P_{23}\right).
	\ee
with $P_{ij}\equiv -2 P_i \cdot P_j$. The integrals may be directly performed by doing a change of variables,
	\be
		t_1=\sqrt{\frac{m_3 m_2}{m_1}}, \quad
		t_2=\sqrt{\frac{m_3 m_1}{m_2}}, \quad
		t_3=\sqrt{\frac{m_1 m_2}{m_3}}. \label{changevars}
	\ee
obtaining
	\be
		A(1,2,3)= \frac{\pi^h}2\, g \,\gt{\sum_i^3 \Delta_i-2h} \mathcal E_3 				\prod_{i=1}^3 \int \frac{\ud m_i}{m_i} m_i^{\delta_{jk}} 
		e^{-m_i P_{jk}}
	\ee
where it should be understood that if $i=1, jk=23$, etc, and
	\be
		\delta_{12}=\frac{\dd1+\dd2-\dd3}2, \quad
		\delta_{23}=\frac{\dd2+\dd3-\dd1}2, \quad
		\delta_{13}=\frac{\dd1+\dd3-\dd2}2.
	\ee
The integrations are now trivial and one obtains
	\be
		A(1,2,3)=\frac{\pi^h}2\,g \, \gt{\sum_i^3 \Delta_i-2h} \mathcal E_3 				\prod_{i<j}^3 \Gamma(\delta_{ij})(P_{ij})^{-\delta_{ij}}
	\ee
In general, we define the normalization constant in \reef{Mellin} by
	\be
	\mathcal N\equiv \frac{\pi^h}2\, \prod_{i=1}^{n}\frac{ C_i}{\Gamma(\Delta_i)}.
	\ee
In this particular case, this gives for the three-point Mellin amplitude:
	\be
		M_3=g\, \gt{\sum_i^3 \Delta_i-2h}\equiv V^{\Delta_1, \Delta_2, \Delta_3}_{[0,0,0]} 
	\ee
The notation for the vertex $V$ will become clear later on. For the practical purpose of computing the Mellin amplitude, we need not worry about the overall normalization constant $\mathcal N$, since to restore it, one can simply include a factor $C_i/\Gamma(\Delta_i)$ for each external leg. As such we will for the most part omit it from our calculations.

\subsection{4-point exchange diagram}
\label{scalarexchange}

\FIGURE[ht]{
	\parbox{10 cm}{
	\centering
\includegraphics[width=7cm]{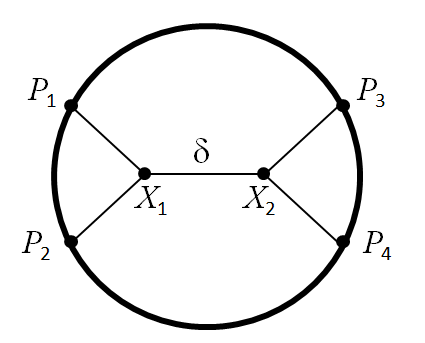}
\caption{Scalar exchange diagram.}
	}
	\label{scexc2}
}

Now let us tackle an example where there is an intermediate state being exchanged in the bulk. We consider a four point amplitude of operators $\mathcal O_i$ and dimension $\Delta_i$, $i=1,\ldots,4$, where a scalar of conformal dimension $\delta$ is being exchanged in the ``s-channel''. The Witten diagram is shown in figure \ref{scexc2}.
Let us denote the corresponding amplitude by $I_s$. There are now two three point interactions happenning at points $X_1, X_2$, over which we must integrate over. The amplitude is written
	\bea
		I_s=g^2\iads \ud X_1 \iads \ud X_2\, E_1(P_1,X_1)E_2(P_2,X_1) 					G_{BB}(X_1,X_2) E_3(P_3,X_2) E_4(P_4,X_2).
	\eea
As we've seen in section \ref{bb props} the dependence of the bulk-to-bulk propagator on $X_1,X_2$ factorises, and the amplitude becomes
	\be
		I_s=\int_{-i \infty}^{+i\infty} \frac{\ud c}{2\pi i} f_{\delta}(c) 
		\ib \ud Q\, A(P_1,P_2,Q_+)
		A(Q_-,P_3,P_4). \label{4pt}
	\ee
with 
	\bea
		A(P_1,P_2,Q_+)=g\, \int_0^{+\infty} 
		\frac{\ud t_1}{t_1} \frac{\ud t_2}{t_2} \frac{\ud s}{s} 
		t_1^{\dd 1} t_2^{\dd 2} s^{h+c} 
		\iads \ud X_1\, e^{2(t_1 P_1+t_2 P_2+s Q)\cdot X_1},\hspace{0.5 cm} \\
		A(P_3,P_4,Q_-)=g\, \int_0^{+\infty} 
		\frac{\ud t_3}{t_3} \frac{\ud t_4}{t_4} \frac{\ud \bar s}{\bar s} 
		t_3^{\dd 3} t_4^{\dd 4} s^{h-c} 
		\iads \ud X_2\, e^{2(t_3 P_3+t_4 P_4+\bar s Q)\cdot X_2} \hspace{0.5 cm}.
	\eea
These are simply three-point amplitudes, which we have already computed. This decomposition is shown diagramatically in figure \ref{decomposition}.
\FIGURE[ht]{
	\parbox{12 cm}{
	\centering
\includegraphics[width=12 cm]{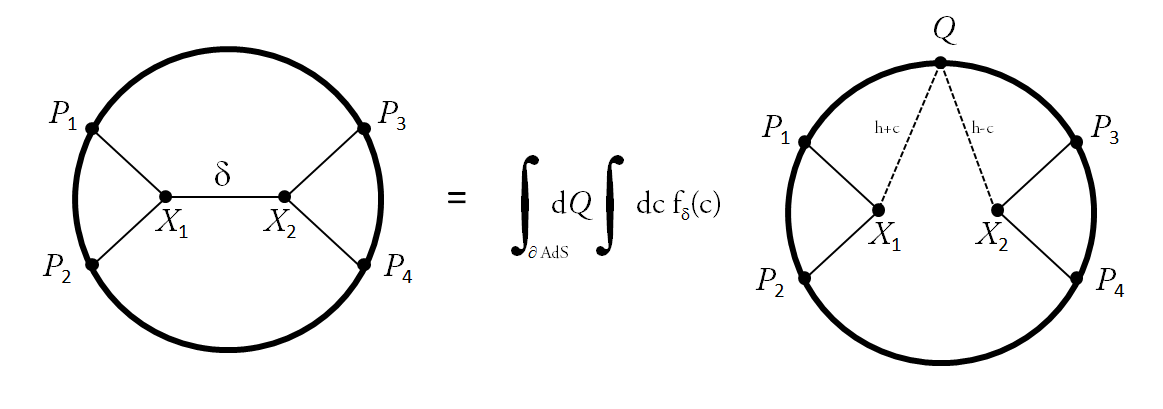}
\caption{Four-point amplitudes result from gluing a pair of three-point amplitudes}
	}
\label{decomposition}
}

Since the bulk-to-bulk propagators always factorise in this way, any $n$-point amplitude will be the result of gluing together several three point amplitudes. We need a useful notation for denoting these, as they will occur often. We choose:
	\be
		A_{\dd i,\dd j,h\pm c_k}(P_i,P_j,Q_i)\equiv A(i,j,c_k^\pm).
	\ee
In case a given three point amplitude contains two $Q's$ then it will also depend on two $c$ parameters. To every boundary coordinate integration there will correspond a single $c$, so that the above notation is consistent. 

To compute integrals such as the one in \reef{4pt}, the standard procedure is to introduce Schwinger parameters to exponentiate the powers of $P_{ij}$. These are the $t$ and $s$ parameters appearing in the expressions for the propagators. In practice, we always start by first performing the $X$ integrations so that we are left with expressions of the form:
	\be
		A(i,j,c^\pm)=g_{i,j,c^\pm}\int_0^{+\infty} 
		\frac{\ud t_1}{t_1} \frac{\ud t_2}{t_2} \frac{\ud s}{s} 
		t_1^{\dd 1} t_2^{\dd 2} s^{h+c} \exp\left[- t_1 t_2 P_{12}+
		2 s (t_1 P_1+t_2 P_2)\cdot Q\right]
	\ee
with
	\be
	g_{i,j,c^\pm}\equiv g\, \pi^{h}\, \gt{\dd i +\dd j+(h\pm c)-2h}.
	\ee
In the particular case at hand, if we write both 3-point amplitudes in this fashion it is easy to see that the $Q$ integral which must be performed is precisely of the form \reef{boundint}. We then get
	\bea
		A_4 &=& g^2 \left(2\pi^{3h}\right)\int_{-i \infty}^{+i\infty} \frac{\ud c}{2\pi i} 				f_{\delta,0}(c) \int \widetilde{\ud^2 s} \,
		\gt{\dd1 +\dd2 +c-h}\gt{\dd3 +\dd4-c-h}\nonumber \\		
		&& \int \prod_{i=1}^4 \frac{\ud t_i}{t_i} t_i^{\dd i}
		\exp\left[-(1+s^2)t_1 t_2 P_{12}-(1+\bar s^2)t_1 t_2 P_{34}
		-s \bar s\,\sum'_{(ij)} t_i t_j P_{ij}\right].
	\eea
where the primed sum indicates we are summing over the ``cross-links'' $13,14,23,24$. We can now use Symanzik's star formula (which we review in appendix \ref{Symanzik}), to show that the amplitude $I_s$ can be written in the form \reef{Mellin}, with a Mellin amplitude given by
	\bea
		M(\delta_{ij})&=& 2
		\int_{-i \infty}^{+i\infty} \frac{\ud c}{2\pi i} 	f_{\delta,0}(c)\,
		I(12,h,c)I(34,h,-c),
	\eea
\vspace{-0.5 cm}
with  e.g.
	\bea
	I(12,h,c)&=&g_{1,2,c^+}\int_0^{+\infty} \frac{\ud s}s 
	s^{h+c-\sum' \delta_{ij}}
	\left(1+s^2\right )^{-\delta_{12}}, \label{I12}
	\eea
The integrals can be evaluated in terms of gamma functions. Using the relations \reef{mandel} to express the $\delta_{ij}$ parameters in terms of Mandelstam invariants we find
	\bea
	M(s_{12})=\frac{g^2 }{\gt{\dd1+\dd2-s_{12}}\gt{\dd3+\dd4-s_{12}}}
	\int_{-i \infty}^{+i\infty} \frac{\ud c}{2\pi i} 						\frac{l_h(c)l_h(-c)}{(\delta-h)^2-c^2} \label{scex1}
	\eea
	where we have defined
	\be
	l_h(c)=
\frac{\gt{h+c-s_{12}}\gt{\dd1+\dd2+c-h}\gt{\dd3+\dd4+c-h}}{2\Gamma(c)}. \label{scex2}
	\ee
The Mellin-Barnes integral can be exactly evaluated in terms of a hypergeometric $_3F_2$ function \cite{Penedones:2010ue}:
	\bea
	&& M(s_{12})=
	\frac 12 \frac{g^2}{s_{12}-\delta}\, \frac{\gt{\Delta_1+\Delta_2+\delta-h} \gt{\Delta_3+\Delta_4+\delta-h}}{\Gamma(1+\delta-h)} 
	\nonumber \\
	&& _3 F_2\left(\frac{2-\Delta_1-\Delta_2+\delta}2,
	\frac{2-\Delta_3-\Delta_4+\delta}2,\frac{\delta-s_{12}}2; \frac{2+\delta-s_{12}}2,1\!+\!\delta-\!h;1\right).
	 \label{intvaluescalar}
	\eea
It is more useful for us however, to write the amplitude in a different fashion. Since the integral must lead to a meromorphic of $s_{12}$, we can write the result as a Laurent series in $s_{12}$.
The poles of this function are found by examining when the $c$ integration contour gets pinched between two poles of the integrand.
We can choose the contour such that this happens when $c=\delta-h$ and $s_{12}=\delta+2n$, with $n$ a positive integer. Then it is easy to find
	\be
	M(s_{12})=\sum_{n=0}^{+\infty} \frac{P_n^\delta}{s_{12}-\delta-2n} 		V_{[0,0,n]}^{\dd1,\dd2,\delta}V_{[0,0,n]}^{\dd3,\dd4,\delta}+\mbox{\ldots}.
	\ee
The dots represent polynomial contributions to the amplitude, but as it happens, in this particular case they are vanishing, as can be checked by computing the amplitude exactly, and the sum of poles is therefore the full amplitude. We have defined the vertices and propagator normalization,
	\bea
	V^{\Delta_1,\Delta_2,\Delta_3}_{[0,0,0]}&=&
		g\, \gt{\sum_i^3 \Delta_i-2h}, \label{v0} \\
	V^{\Delta_1,\Delta_2,\Delta_3}_{[0,0,n_1]}&=&
		V^{\Delta_1,\Delta_2,\Delta_3}_{[0,0,0]} 
		\left(1-\frac 12 \sum_i^3\Delta_i+\Delta_3\right)_{n_1} \label{v1} \\
	P_n^\delta&=& \left[2\,n!\, \Gamma\left(1+\delta-h+n\right)\right]^{-1} 			\label{propnorm}	
	\eea
with the help of the Pochhammer symbol $(a)_m=\Gamma(a+m)/\Gamma(a)$. The interpretation of this expression is clear: the Mellin amplitude is an infinite sum of products of three point vertices and a propagator. The sum runs over the propagating fields, which include a field with conformal dimension $\delta$ and its ``descendants'', with dimension $\delta+2n$. From the above one reads off the three point Mellin amplitude of two fields of dimensions $\dd 1, \dd2$, and one such descendant to be simply $V_{[0,0,n]}^{\dd3,\dd4,\delta}$. In particular for $n=0$ this reduces to the three point Mellin amplitude we previously computed.

This result suggests a set of Feynman rules for Mellin amplitudes, where to each internal line in a Witten diagram one associates an infinite sum of propagating fields (one primary and an infinite set of descendants), to each vertex one associates a factor $V_{[m,n,p]}^{\dd1,\dd2,\dd3}$, and for each line a normalization factor which is  the inverse of $\Gamma(1+\Delta_i+n-h)$. These are of course nothing but the Feynman rules we conjectured in the introduction section. However, right now we do not yet know the form of the general vertex, which can involve up to three ``descendants''.
In principle its form is directly fixed by kinematic considerations alone, that is, by conformal symmetry. In practice,  to proceed we shall extract this vertex by evaluating higher point amplitudes. This provides a simple way of reading off the vertex, and will also act as a cross-check on our proposed Feynman rules.

Firstly we consider a five point amplitude. In such a diagram there is a vertex connecting two internal lines, and from it we will be able to read off $V_{[0,n,p]}^{\dd1,\dd2,\dd3}$. We will also explicitly see that these Feynman rules still work there. Finally, the full vertex may be obtained by considering a 6-point amplitude. We shall see how the latter can be written as a product of three propagators and associated vertices, and read off $V_{[m,n,p]}^{\dd1,\dd2,\dd3}$.

\section{Scalar higher-point amplitudes}
\subsection{5-point amplitude}

Consider the Witten diagram of figure \ref{fivept}, for a five point amplitude in cubic theory.
\FIGURE[ht]{
	\parbox{10 cm}{
	\centering
	\includegraphics[width=6cm]{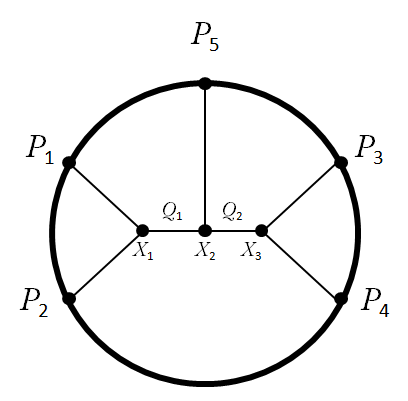}
	\caption{A five-point Witten diagram in scalar theory}.
	
	}
\label{fivept}
}
The amplitude is given by\footnote{For economy of space we omit the external line normalization factors $C_i/\Gamma(\Delta_i)$, which are removed anyway upon passage to Mellin space.}
	\bea
		A_5= g^3  \int \prod_{i}^5 \frac{\ud t_i}{t_i} 				t_i^{\Delta_i} 
		\int_{-i\infty}^{+i \infty} \frac{\ud c_1 \ud c_2}{(2\pi i)^2} f_{\delta_1}(c_1)f_{\delta_2}(c_2)\int \widetilde{\ud^2 s_1}\widetilde{\ud^2 s_2}&& \nonumber \\
	\ib 	\ud Q_1 \ud Q_2  \iads \ud X_1 \ud X_2 \ud X_3  
	\exp \bigg[2 X_1\cdot(t_1 P_1+t_2 P_2+s_1 Q_1)+&&\nonumber \\
	 2 X_3\cdot(t_3 P_3+t_4 P_4+s_2 Q_2)+2 X_2\cdot(t_5 P_5+\bar s_1 Q_1+\bar s_2 Q_2)\bigg].&& \nonumber
\eea
This looks quite complicated as it stands. However, we see that as expected from our general arguments in the previous section, each $X_i$ only couples to three coordinates coming into a vertex, and so we can immediately write
	\bea
	A_5=g^3 \int_{-i\infty}^{+i \infty} \frac{\ud c_1 \ud c_2}{(2\pi i)^2}
	f_{\delta_1}(c_1)f_{\delta_2}(c_2)\ib \ud Q_1\, \ud Q_2\,  				A(1,2,c_1^+)A(3,4,c_2^+)A(5,c_1^-,c_2^-)
	\eea
Replacing the three point amplitudes for their Schwinger-parameterized expressions, we have an integrand of the form
	\bea
	\simeq	\exp \bigg[-t_1 t_2 P_{12}-t_3 t_4 P_{34}\bigg] \exp\bigg[2 		Q_1\cdot (s_1 t_1 P_1+s_1 t_2 P_2+\bar s_1 t_5 P_5) + \nonumber \\
		2 Q_2\cdot (s_2 t_3 P_3+s_2 t_4 P_4+ \bar s_2 t_5 P_5)+2 \bar s_1 		\bar s_2 Q_1\cdot Q_2\bigg]
	\eea
We now perform the $Q$ integrals, first $Q_1$ and $Q_2$. Consequently the result appears to break the symmetry of the diagram, but this will be restored later. The result is that the integrand becomes the exponential of a polynomial quadratic in the $P_i$'s of the form
	\be
	\simeq	\exp\bigg[-\sum_{i<j} Q_{ij} t_i t_j P_{ij}\bigg]
	\ee
Using Symanzik's star formula we obtain the Mellin amplitude
	\bea
	&& M_5=g^3 \left(4\pi^h\right)\int_{-i\infty}^{+i\infty} \frac{\ud c_1 \ud c_2}{(2\pi i)^2}f_{\delta_1}(c_1)f_{\delta_2}(c_2) %
	\,g_{1,2,c_1^+}\,g_{3,4,c_2^+}\,g_{5,c_1^-,c_2^-} \nonumber \\
	&& \int \widetilde{\ud^2 s_1}\widetilde{\ud^2 s_2} 
	\left(1+s_1^2+s_1^2 \bar s_1^2 \bar s_2^2\right)^{-\delta_{12}}(1+s_2^2)^{-\delta_{34}}\left(1+\bar {s_1^2}\right)^{-\delta_{35}-\delta_{45}}
\times\nonumber \\
&& \left(1+\bs2^2+\bs1^2\bs2^2\right)^{-\de15-\de25} \left(s_1 \bs1\right)^{-\de15-\de25-\de13-\de14-\de23-\de24}\,\left(s_2 \bs2\right)^{-\de35-\de45-\de13-\de14-\de23-\de24}.
	\eea
To proceed we must compute the $s_i$ integrals. The integrals of $s_1, s_2$ are simply performed and result in more Gamma functions. Using the Mandelstam invariant representation of the $\delta_{ij}$, the amplitude becomes
\bea
M_5= 
	g^3 \pi^{4h}
 \int_{-i\infty}^{+i\infty} \frac{\ud c_1 \ud c_2}{(2\pi i)^2} f_{\delta_1}(c_1) f_{\delta_2}(c_2) \frac{\gt{-s_{12}+c_1+h}\gt{-s_{34}\!+\!c_2\!+\!h}}{
	\gt{\dd1\!+\!\dd2\!-\!s_{12}}
	\gt{\dd3+\dd4-s_{34}}}
\gt{\dd5\!-c_1\!-c_2}\nonumber \\
\gt{\dd3\!+\!\dd4\!+\!c_2\!-\!h} 
\gt{\dd1\!+\!\dd2\!+\!c_1\!-\!h}
\gt{\dd3\!+\!\dd4\!-\!c_2\!-\!h} 
\gt{\dd1\!+\!\dd2\!-\!c_1\!-\!h}
&& \nonumber \\
\int\frac{\ud \bs1}{\bs1}\frac{\ud \bs2}{\bs2}
 \bs1^{h-c_1-s_{12}}\bs2^{h-c_2-s_{34}}
\!\left(1+\bs1^2\right)^{\!\!-\de35-\de45}\left(1+\bs2^2+\bs1^2\bs2^2\right)^{\!\! -\de15-\de25}\!\!\left(1+\bs1^2\bs2^2\right)^{\frac{s_{12}-c_1-h}2}. && \label{intfive}
\eea
Let us focus on the integral on the third line. After a change of variables into $x=\bs1^2, y=\bs2^2$ the integral becomes of the form
\be
\simeq \int_{0}^{+\infty} \int_{0}^{+\infty} \frac{\ud x}x\, \frac{\ud y}y \, x^{a} y^{b} (1+x)^{c}(1+y+x y)^{d}(1+x y)^{e},
\ee
with 
\bea
a=\frac{h-c_1-s_{12}}2, \qquad b=\frac{h-c_2-s_{34}}{2}, \qquad c=-(\de35+\de45), \nonumber \\  d=-\de15-\de25, \qquad e=\frac{s_{12}-c_1-h}2
\eea
This integral possesses a large number of symmetries interchanging the exponents of the various factors. Rescaling $y\to y/(1+x)$ followed by $x\to 1/x$ we obtain
\be
\int_{0}^{+\infty} \int_{0}^{+\infty} \frac{\ud x}x\, \frac{\ud y}y \, x^{-a+b-c} y^b (1+x)^{-b+c-e}(1+y)^d(1+x+y)^e \label{5ptnice}
\ee
To compute this integral, first perform the change of variables
\be
x\to \frac{x}{1-x}, \qquad y\to \frac{y}{1-y}
\ee
whereupon the integral becomes
\bea
&& \int_{0}^{1} \int_{0}^{1} \frac{\ud x}x\, \frac{\ud y}y \, x^{-a+b-c} y^b (1-x)^{-1+a}(1-y)^{-1-b-d-e}(1-x y)^e \nonumber \\
&\equiv&
\int_{0}^{1} \int_{0}^{1} \frac{\ud x}x\, \frac{\ud y}y \, x^{a_1} y^{a_2} (1-x)^{-a_1+b_1-1}(1-y)^{-a_2+b_2-1}(1-x y)^{a_3}.
\eea
The integral can be performed assuming $\mbox{Re}(b_k)>\mbox{Re}(a_k)>0$, for $k=1,2$ using
\bea
\ _3F_2(a_1,a_2,a_3;b_1,b_2;z)=\hspace{10 cm}&&  \nonumber \\
\prod_{k=1}^2 \frac{\Gamma(b_k)}{\Gamma(a_k)\Gamma(b_k-a_k)}\int_{0}^{1}\!\! \int_{0}^{1} \frac{\ud x}x\, \frac{\ud y}y \, x^{a_1} y^{a_2} (1-x)^{-a_1+b_1-1}(1-y)^{-a_2+b_2-1}(1-z x y)^{-a_3}&&  \ \ \ \ \ \ \label{3f2}
\eea
and so we obtain
\bea
\int\int (\ldots)=
\gt{c_1\!-\!c_2\!+\!\dd5} \gt{c_1\!+\!c_2\!+\!\dd5}\gt{\!-c_1\!+\!h\!-\!s_{12}}\gt{-c_2\!+\!h\!-\!s_{34}}  &&\nonumber\\
\times \ \frac{ _3F_2\left(\{\frac{c_1\!-\!c_2\!+\!\dd5}{2},\frac{-c_2\!+\!h\!-\!s_{34}\!}2,\frac{c_1\!+\!h\!-\!s_{12}}2,\{ \frac{-c_2\!+\!\dd5\!+\!h\!-\!s_{12}}2 \},\frac{c_1\!+\!\dd5\!+\!h\!-\!s_{34}}2\};1\right)}
{\gt{-c_2+\dd5+h-s_{12}}\gt{c_1+\dd5+h-s_{34}}}
. &&  
\eea
The $ _3F_2$ hypergeometric function at argument $z=1$ satisfies a number of identities, among which
\bea
_3F_2(a_1,a_2,a_3,b_1,b_2;1)=\frac{\gn{b_1}\gn{b_2}\gn{b_1+b_2-a_1-a_2-a_3}}{\gn{a_1}\gn{b_1+b_2-a_1-a_2}\gn{b_1+b_2-a_1-a_3}} && \nonumber \\
\ _3 F_2(b_1-a_1,b_2-a_1,b_1+b_2-a_1-a_2-a_3,b_1+b_2-a_1-a_2,b_1+b_2-a_1-a_3;1) && \ \ \ 
 \eea
which exchanges the roles of $c_1, s_{12}$ with $c_2, s_{34}$.

The expression for the Mellin amplitude is then written as
%
\bea
&& M_5= \frac{g^3}{\gt{\dd1+\dd2-s_{12}}\gt{\dd3+\dd4-s_{34}}} \int_{-i\infty}^{+i\infty} 
	\frac{\ud c_1 \ud c_2}{(2\pi i)^2}\,
	\frac{L_{1}(c_1)L_{1}(-c_1)}{(\delta_1-h)^2-c_1^2}
	\frac{L_{2}(c_2)L_{2}(-c_2)}{(\delta_2-h)^2-c_2^2}
	\,  \nonumber \\
&&
\prod_{\sigma_1,\sigma_2=\pm 1}\!\! \! \gt{\sigma_1 c_1\!+\!\sigma_2 c_2\!+\!\dd5}\!\!
 \frac{ 
 _3F_2\left(\{ 
 	\frac{c_1\!-\!c_2\!+\!\dd5}{2},
	\frac{-c_2\!+\!h\!-\!s_{34}\!}2,
	\frac{c_1\!+\!h\!-\!s_{12}}2\},
	\{
\frac{-c_2\!+\!\dd5\!+\!h\!-\!s_{12}}2,
\frac{c_1\!+\!\dd5\!+\!h\!-\!s_{34}}2\},
1\right)
}
	{\gt{-c_1+c_2+\dd5}\gt{-c_2+\dd 5+h-s_{12}}\gt{c_1+\dd 5+h-s_{34}}}, \ \ \ \ \ \ \ 
\eea
with
\bea
L_{1}(c_1)=\frac{\gt{c_1+h-s_{12}}\gt{\dd1+\dd2+c_1-h}}{2\gn{c_1}},\quad L_{2}(c_2)=\frac{\gt{c_2+h-s_{34}}\gt{\dd3+\dd4+c_2-h}}{2\gn{c_2}}. \label{Lij}
\eea
%
%
%
The identity between these two expressoins will be shown in the next section. We are interested in obtaining the poles and respective residues in $s_{12}$ and $s_{34}$ of the expression above. Although there are various sets of poles in $c_1$ and $c_2$, the only ones which will end up giving expressions containing poles in $s_{12}$ and $s_{34}$ are the ones at $c_1=\delta_1-h$, $c_2=\delta_2-h$. Computing the residues at these poles we find
\be
M_5=\sum_{n_1,n_2=0}^{+\infty}
\frac{P_{n_1}^{\delta_1}}{s_{12}-\delta_1-2n_1}
\frac{P_{n_2}^{\delta_2}}{s_{34}-\delta_2-2n_2}
\, V_{[0,0,n_1]}^{\dd1,\dd2,\delta_1}
V_{[0,0,n_2]}^{\dd3,\dd4,\delta_2}
V_{[0,n_1,n_2]}^{\dd5,\delta_1,\delta_2}+\ldots \label{m5}
\ee
where the dots represent possible subleading contributions. The only new ingredient in the above is
\bea
V_{[0,n_1,n_2]}^{\Delta_1,\Delta_2,\Delta_3}
=g \,\gt{\sum_i \Delta_i-2h}\,
\left(1-\frac 12 \sum_i^3\Delta_i+\Delta_2\right)_{n_1}
\left(1-\frac 12 \sum_i^3\Delta_i+\Delta_3\right)_{n_2}
\times &&\nonumber \\ \ _3F_2\left(\left\{\frac{\sum_i\Delta_i-2h}2,-n_1,-n_2\right\},\left\{ \frac{\sum_i\Delta_i-2\Delta_2-2n_1}2,\frac{\sum_i\Delta_i-2\Delta_3-2n_2}2\right\},1\right)&& \label{doublevertex}
\eea
It is easy to check that when one or more of the $n_i$'s vanish we reproduce our previous expressions \reef{v0},\reef{v1}.
It's been a long way, but the final result \reef{m5} is particularly simple, and it agrees with the Feynman rules we have defined previously, assuming that the subleading contributions in the above vanish. Attempts to evaluate the Mellin amplitude numerically suggest this is the case, although further work is necessary.
The upshot of this calculation is that we have now in our possession a further ingredient for such rules, which is the vertex for the case where we have two ``descendant'' fields and one primary.

\subsection{6-point amplitude}

The next step is to calculate a six point diagram involving three bulk-to-bulk propagators connected at a single vertex in order to obtain $V^{\dd1,\dd2, \dd3}_{[n_1,n_2,n_3]}$. With this purpose in mind we now turn our attention to the particular Witten diagram in figure \ref{6pt}.
\FIGURE[ht]{
	\parbox{10 cm}{
	\centering
		\includegraphics[width=7 cm]{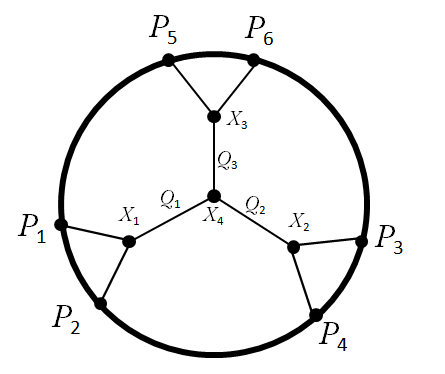}
		\caption{A six-point Witten diagram in scalar theory.}

	}
	\label{6pt}
}
We can immediately write
	\bea
	A_6= \int_{-i\infty}^{+i\infty}\prod_{k=1}^{3} 
	\frac{\ud c_k}{2\pi i}f_{\delta_k}(c_k)\ib \prod_{i=1}^3\ud Q_i \, 
	A(1,2,c_1^+)A(3,4,c_2^+)A(5,6,c_3^+)A(c_1^-,c_2^-,c_3^-)\hspace{1 cm} &&
	\eea

 %
 %
The calculation proceeds as for the five point amplitude - we integrate over each $Q_i$ in turn. Exactly as before one can use the Symanzik star formula to read off the Mellin amplitude. After performing the $s_1, s_2, s_3$ integrals (just like before we could immediately do the integrals in $s_1$ and $s_2$), we are still left with a seemingly complicated integral in $\bar s_1, \bar s_2, \bar s_3$, analogous to the second line of \reef{intfive}. However, as we show in appendix \ref{details}, performing a change of variables it is possible to write the Mellin amplitude as
%
%
%
\bea
	M_6=\frac{g^4}{2^6}\int_{-i \infty}^{i \infty} \prod_{i=1}^{3}
	\left(
	\frac{\ud c_i}{2\pi i}\, 								\frac{
	\gt{\Delta_{i,1}+\Delta_{i,2}+c_i-h}
	\gt{\Delta_{i,1}+\Delta_{i,2}-c_i-h}
	\gt{c_i+h-s_i}
	}
	{\gt{\Delta_{i,1}+\Delta_{i,2}-s_i} \Gamma(c_i)\Gamma(-c_i)\left[(\delta_i-h)^2-c_i^2\right]}\right)\hspace{0.5 cm} &&
\nonumber \\
\gt{h\!-\!c_1\!-\!c_2\!-\!c_3}\int_{0}^{+\infty}\! 
\frac{\ud x}{x}\frac{\ud y}{y}\frac{\ud z}{z}
x^a y^b z^c (1+x)^d(1+y)^e(1+z)^f(1+x+y+z)^g, &&\label{m6} 
\eea
with $\Delta_{i,j}$ the dimension of the $j$th field of the $i$th pair of legs - $j=1,2$ and $i=1,2,3$. For instance, $\Delta_{2,1}\equiv\Delta_3, \Delta_{3,2}\equiv \Delta_6,\ldots$. Also, the $s_i$ variables are the Mandelstam variables associated with each pair of legs, such that $s_1\equiv s_{12}, s_2\equiv s_{34}$ and $s_3\equiv s_{56}$. As for the parameters $a,b,\ldots, g$ we have $g=\frac 12 (c_1+c_2+c_3-h)$ and
	\bea
	a=\frac 12 (-c_1+h-s_{12}), \quad 
	b=\frac 12 (-c_2+h-s_{34}) \quad
	c=\frac 12 ( -c_3+h-s_{56}) &&\nonumber \\
	d=\frac 12 (-c_1-h+s_{12}), \quad
	e=\frac 12 (-c_2-h+s_{34}), \quad
	f=\frac 12 (-c_3-h+s_{56}),&&
	\eea
To proceed we must evaluate the integral on the second line of \reef{m6}. First we do a multinomial expansion on the last factor of the integrand,
	\bea
	(1\!+\!x\!+\!y\!+\!z)^g=\!\!\!\sum_{m_1,m_2,m_3=0}^{+\infty} 						\!\!(-g)_{m_1}(-g\!+\!m_1)_{m_2}(-g\!+\!m_1\!+\!m_2)_{m_3}
	\frac{(-x)^{m_1}}{m_1!}\frac{(-y)^{m_2}}{m_2!}\frac{(-z)^{m_3}}{m_3!}
	\hspace{0.5 cm}&&
	\eea
We are then free to perform the separate integrations over $x,y,z$. The result is
	\bea
	\int_{0}^{+\infty}\! \int_{0}^{+\infty}\!\int_{0}^{+\infty}(\ldots)= 		\prod_{i=1}^3 \frac{\gn{c_i}\gt{-c_i+h-s_i}}{\gt{c_i+h-s_i}} \, 			F_A^{(3)}\left(-g,\left\{a,b,c\right\},\left\{d,e,f\right\};1,1,1\right)
	\eea
where $s_1=s_{12},\ldots$ and $F_A^{(3)}$ is a Lauricella generalized hypergeometric function of three variables \cite{Lauricella, Srivastava, mathworld}. For future reference we give the definition of the Lauricella function $F_A^{(m)}$:
\bea
F_A^{(m)}\left(g,\left\{a_1,\ldots,a_m\right\},\left\{b_1,\ldots,b_m\right\};x_1,\ldots,x_m\right)\equiv 
\sum_{n_i=0}^{+\infty}\left( (g)_{\sum_{i=1}^m n_i} \prod_{i=1}^m\frac{ (a_i)_{n_i}}{(b_i)_{n_i}} \frac{x_i^{n_i}}{n_i!}\right) \label{lauricella}
\eea
The above series is convergent only for  $\sum_i |x_i|<1$. Our interpretation then is to define the sum at this point as the value of the Lauricella function at that point, which is well defined via analytic continuation. Of course it might very well happen that for specific values of the parameters $g,a_i,b_i$ the series reduces to a sum, in which case everything is perfectly well defined.
		
The Mellin amplitude is exactly given by
	\bea
	M_6&=&g^4\int_{-i \infty}^{i \infty} \prod_{i=1}^{3}
	\left(
	\frac{\ud c_k}{2\pi i}\, 								\frac{L_i(c_i)L_i(-c_i)}{(\delta_i-h)^2-c_i^2}\right)
	 \gt{h-c_1-c_2-c_3}
	\nonumber \\
	&& \prod_{i=1}^3 \left(\frac{\Gamma(c_i)}{\gt{\Delta_{i,1}+\Delta_{i,2}-s_i}\gt{c_i+h-s_i}} \right)
F_A^{(3)}\left(-g,\left\{a,b,c\right\},\left\{d,e,f\right\};1,1,1\right).
	\hspace{0.5 cm}
	\eea
with the $L_i$ defined analogously to \reef{Lij}. 
Evaluating the integral above in closed form seems like a difficult challenge. The poles in $s_{12}, s_{34}, s_{56}$ however, are easily found by pinching of two poles in the $c_1,c_2$ and $c_3$ integrations respectively, using the definition \reef{lauricella} of $F_A^{(3)}$. The end result is the remarkably simple expression
\be
M_6=\sum_{n_1,n_2,n_3=0}^{+\infty} \left(\prod_{i=1}^{3} \frac{P_{n_i}^{\delta_i}}{s_i-\delta_i-2\,n_i}\right) 
V_{[0,0,n_1]}^{\dd1,\dd2,\delta_1}
V_{[0,0,n_2]}^{\dd3,\dd4,\delta_2}
V_{[0,0,n_3]}^{\dd5,\dd6,\delta_3}
V_{[n_1,n_2,n_3]}^{\delta_1,\delta_2,\delta_3}+\ldots
\ee
This not only provides further evidence for our set of Feynman rules for Mellin amplitudes, but also gives us the final vertex
\bea
V_{[n_1,n_2,n_3]}^{\dd1,\dd2,\dd3}
= V_{[0,0,0]}^{\dd1,\dd2,\dd3}
\left(1-h+\dd1\right)_{n_1}\left(1-h+\dd2\right)_{n_2}\left(1-h+\dd3\right)_{n_3} \hspace{1 cm} &&\nonumber \\
\ \ F_A^{(3)}\left(\frac {\dd1\!+\!\dd2\!+\!\dd3\!-\!2h}2,\left\{-\!n_1,-\!n_2,-\!n_3\right\},\left\{1\!+\!\dd1\!-\!h,1\!+\!\dd2\!-\!h,1\!+\!\dd3\!-\!h\right\};1,1,1\right).\hspace{0.5 cm}&&   \label{3ptvertex}
\eea
Notice that with $n_i$ positive integers, the Lauricella triple hypergeometric function is given by a finite sum. 

Now let us show that the vertex function $V_{[n_1,n_2,n_3]}^{\dd1,\dd2,\dd3}$ just computed reduces to the previous expression \reef{doublevertex} when one of the integers $n_i$ is zero. When this happens, one of the sums in the definition \reef{lauricella} reduces to a single term, and the Lauricella triple hypergeometric function reduces to the Appell $F_2$ function, which we denote by $F_A^{(2)}$. For instance, if $n_3=0$ we get
\bea
&&V_{[n_1,n_2,n_3]}^{\dd1,\dd2,\dd3}
= V_{[0,0,0]}^{\dd1,\dd2,\dd3}
\left(1-h+\dd1\right)_{n_1}\left(1-h+\dd2\right)_{n_2}\ \ \ \  \nonumber \\
&& F_A^{(2)}\left(\frac {\dd1\!+\!\dd2\!+\!\dd3\!-\!2h}2,\left\{-\!n_1,-\!n_2\right\},\left\{1\!+\!\dd1\!-\!h,1\!+\!\dd2\!-\!h\right\};1,1\right).\ \ \  \label{doublevertex2}
\eea
The Appell $F_2$ function with arguments $x=y=1$ is directly related to the $ _3F_2$ hypergeometric function at argument $x=1$. In order to prove this one computes the integral
\be
\int_{0}^{+\infty} \int_{0}^{+\infty} \frac{\ud x}x\, \frac{\ud y}y \, x^{-a+b-c} y^b (1+x)^{-b+c-e}(1+y)^d(1+x+y)^e
\ee
in two different ways, firstly by using formula \reef{3f2}, and secondly using the multinomial expansion on $(1+x+y)^e$ and integrating. The end result is
\bea
F_A^{(2)}\left(e,\left\{a,b\right\},\left\{c,d\right\}; 1,1\right)= \hspace{8 cm}&&\nonumber \\
\frac{\poc{1+a-c+e}{-a}\poc{1+b-d+e}{-b}}{\poc{1+a-c}{-a}\poc{1+b-d}{-b}} \ _3F_2\left(\{a,b,e\},\{1+a-c+e,1+b-d+e\},1\right)&& \ \ \ 
\eea
Using this identity it is straightforward to show that \reef{doublevertex2} reduces to \reef{doublevertex}.

\subsection{Outline of the 12-point amplitude calculation}

The beautiful expression \reef{3ptvertex} for the general vertex $V_{[n_1,n_2,n_3]}^{\dd1,\dd2,\dd3}$ in $\phi^3$ theory leads us to conjecture that in $\phi^{m}$ theory the general vertex takes the form given in the introduction,
\bea
 && V^{\dd 1\ldots \dd m}_{[n_1,\ldots, n_m]}=g_m \, \gt{\sum_i \dd i-2h}
	\left(\prod_{i=1}^m 	\left(1-h+\dd i\right)_{n_i}\right)\nonumber \\
 && F_A^{(m)}\left(\frac {\sum_{i=1}^n \dd i-\!2h}2,\left\{-\!n_1,\ldots,-\!n_m\right\},\left\{1\!+\!\dd1\!-\!h,\ldots,1\!+\!\dd m\!-\!h\right\};1,\ldots,1\right). \hspace{1 cm} \ \ \  
	\eea
As a rather non-trivial check of this, we have performed the computation of a twelve-point amplitude in $\phi^4$ theory. The calculation is tedious but essentially the same as in the six point function in $\phi^3$ theory. The diagram is of the form given in figure \ref{12pt}.
\FIGURE[ht]{
	\parbox{10 cm}{
	\centering
\includegraphics[width=7cm]{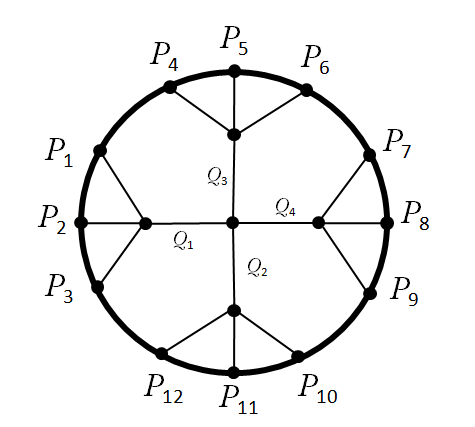}
\caption{A twelve-point diagram in $\phi^4$ theory.}

	}
\label{12pt}
}
The computation of this amplitude is very similar to the six-point calculation. The $X$ integrals are performed trivially as usual. The integrals over boundary coordinates $Q_i$ are also trivial, and the resulting expression can be translated into a Mellin amplitude consisting of four $c_i$ integrals, and a set of four $s_i, \bar s_i$ integrals. The latter can be explicitly performed, while the former lead to poles in the various Mandelstam variables upon pinching. The $s_i$ integrals can be carried out immediately as in the four-, five- and six-point amplitude calculations, so that the only non-trivial part of the calculations are the remaining integrals over the $\bar s_i$ parameters. At this point we are in a situation similar to that described in appendix \reef{details}, with a rather nasty looking integrand. However, by performing change of variables of the type described in that same appendix, the integral can be successively simplified until it reduces to
\bea
	\int_{0}^{+\infty}\prod_{i=1}^4\left[\ud x_i\, 
x_i^{a_i} (1+x)^{b_i}  \right ] \left(1+\sum_i^4 x_i\right)^g\ , && 
\eea
with $g=\frac 12 (c_1+c_2+c_3+c_4-h)$,
	\bea
	a_i=\frac 12(-c_i+h-m_i), \quad b_i=\frac 12(-c_i-h+m_i)
	\eea
and the four Mandelstam variables $m_i$ are $m_1\equiv s_{123}, m_2\equiv s_{456},\ldots$. To evaluate the integral we perform a multinomial expansion as before, which leads to the four-variable Lauricella function. The calculation then proceeds as for the six-point function and one precisely finds an expression for the poles of the Mellin amplitude consistent with the Feynman rules conjectured in the introduction.

\section{Conformal invariance of index structure}
\label{confinv}

In the following sections we will be interested in evaluating amplitudes which involve fields carrying spin degress of freedom, either in an internal propagator or as an external state. In the latter case, to obtain expressions for amplitudes in $d$-dimensional space, we will have to contract the $M$ indices with the pull-backs $\zeta_\mu^M$. These in turn are contracted with some polarization tensors, so that overall we may say that the $M$ indices are contracted with polarizations $\xi^M$. These polarizations satisfy
	\be
	\xi_1 \cdot P_1=\xi_1^{M_1} P_{1,M_1}=\xi^\mu \frac{\partial P_1^{M_1}}{\partial y_1^\mu} \, P_{1,M_1}=0
     \ee
because of the condition $P_1^2=0$. Further, we have
	\be
	P_M D_{\Delta}^{MA}=P^A(1+ \frac{1}{\Delta} P^M \partial_M)=0
     \ee
The rightmost factor checks that the overall amplitude scales with $P$ like $1/P^\Delta$, which has to be the case, and so it is vanishing. That is, the transversality condition of $J^{MA}$ has transformed into a scaling condition imposed by $D^{MA}$. In this way, $D^{MA}$ can be thought of as a projector which implements conformal symmetry of the index structure.

Overall, these results are very suggestive: in the embedding formalism, amplitudes depend on objects $P_i$ and polarizations $\xi_i$ such that
	\be
	P_i^2=0, \qquad \xi_i\cdot P_i=0, \qquad \xi_i \simeq \xi_i+P_i. \label{gaugecond}
	\ee
These are exactly the conditions required of a gauge theory amplitude depending on momenta $P_i$ and polarizations $\xi_i$. This suggests that $d$-dimensional CFT dynamics are related to gauge (or gravity) theories in $d+2$ dimensions, but where the coordinates of the one are related to the momenta of the other. Although we will not try to flesh out this relation further here, the above set of requirements above already imply strong constraints on the possible index structure of conformally invariant amplitudes. 

Consider for instance an amplitude of the form $\left \langle J^M_3(P_3) \mo (P_1)\mo (P_2)\right \rangle$. On the one hand, no $P_3$ with free indices are allowed, so that the index dependence must be carried by $P_1$, $P_2$. Then ``gauge invariance'' uniquely fixes the structure
	\be
	\frac{P_{1}^{M_3}}{P_{13}}-\frac{P_{2}^{M_3}}{P_{23}}
	\equiv X_{\mbox{\tiny{12}}}^{M_3}.
	\ee
The rest of the amplitude is fixed by requiring the correct behaviour under rescalings of $P_1, P_2, P_3$ by constant factors. Generically, the only structures which can appear in any amplitude are of the form above or
	\be
	I^{M_1 M_2}\equiv \eta^{M_1 M_2}-\frac{P_1^{M_2} P_2^{M_1}}{P_1\cdot P_2}
	\ee
which vanishes upon contraction with either $P_1^{M_1}$ or $P_2^{M_2}$. 

In particular, consider a current three-point function. The general structure of such amplitudes, as imposed by conformal invariance has been known for a long time. With our methods, finding the index structure of such an amplitude is a trivial task: there are only two possible structures, namely
	\be
	X_{\mbox{\tiny{23}}}^{M_1}X_{\mbox{\tiny{13}}}^{M_2}
	X_{\mbox{\tiny{12}}}^{M_3}, \qquad \mbox{or} \qquad
	I^{M_1 M_2} X_{\mbox{\tiny{12}}}^{M_3}+\mbox{permutations}
	\ee
And indeed, this is correct. A similar argument can be made for the four-point function. All terms are of the form
\be
I\,I, \qquad I X X, \qquad X X X X,
\ee
but there are a greater number of them, as one could have several $X's$ with the same index, {\em e.g.} $X_{\mbox{\tiny{12}}}^{M_4}$,$X_{\mbox{\tiny{13}}}^{M_4}$, $X_{\mbox{\tiny{23}}}^{M_4}$. 

The arguments given above are completely general, in the sense that they apply to any conformal correlation function independently of the spin or number of fields involved. In other words, the most general amplitude must have an index structure such that it reduces to polynomials in $I$, $X$. In general,  current conservation places constraints on the final form of the amplitude by relating the coefficients of different kinds of index structures. However such constraints do not seem to have a simple formulation in the embedding formalism, and they are most usefully seen by pulling back our expressions to $d$-dimensions.

Actually there is a slight subtlety we have ommitted. It is easiest to see the problem in the case of the stress-tensor. This is the question of removal of traces from the index structure, which can be understood by the simple example of the correlator of a stress-tensor and two scalar fields. The index structure of such a correlator is completely fixed by conformal invariance\footnote{See for instance \cite{Osborn:1993cr}}, and we get
	\be
	\left \langle T^{M_3 N_3}(P_3)\mo (P_1)\mo (P_2)\right \rangle
	\propto X_{12}^{M_3} X_{12}^{N_3}-\mbox{trace}.
	\ee
The question is, what exactly do we mean by the trace part removal in the above? If we remove the $(d+2)$ dimensional trace of the expression above, so that it becomes
	\be
	\simeq X_{12}^{M_3} X_{12}^{N_3}-\frac{1}{d+2} \eta^{M_3 N_3} 		(X_{12})^2,
	\ee
then we lose ``gauge invariance'' as easily seen. Another problem is that it is $1/(d+2)$ which appears in the expression, whereas we expect the final result to be traceless in $d$ dimensions, not $d+2$. As it turns out, both these problems can be solved at once. To restore gauge invariance we must, counter-intuitively, add gauge-variant terms. To do this, first introduce introduce the vector $Q$ which in our parameterization is simply $Q=(Q^+,Q^-,Q_{\mu})=(0,1,0)$. This implies that $Q\cdot P=-1/2$, for any boundary point $P$. Then to remove the trace we take
	\be
	\simeq X_{12}^{M_3} X_{12}^{N_3}-\frac{1}{d}\left (\eta^{M_3 N_3}+4 P^{(M_3} Q^{N_3)}\right) (X_{12})^2,
	\ee
It is easily checked that the expression above is both gauge-invariant and traceless, at least in $d+2$ dimensions. Also, the extra terms we have introduced vanish upon contraction with the pull-backs $\zeta_\mu^M$, and we obtain an expression which is traceless in $d$ dimensions. 

This result is more general, and it applies to any pair of symmetric traceless indices. It follows from \cite{Weinberg:2010fx}
	\be
	\eta^{\mu\nu}\zeta_{\mu}^{M_1}(P)\zeta_{\nu}^{M_2}(P) T_{M_1 M_2\ldots}(P)=\eta^{M_1 M_2} T_{M_1 M_2\ldots}(P), \quad \mbox{if} \quad  P^M_1 T_{M_1\ldots }(P)=0
	\ee
which is easily proved noting that in our parameterization we have
	\be
	\eta^{\mu\nu}\zeta_{\mu}^{M_1}\zeta_{\nu}^{M_2}= 				\left(\eta_{M_1 M_2}+4 Q^{(M_1} P^{M_2)} \right).
	\ee
With these results, we may say that before taking traces, the amplitude is indeed fully written in terms of the objects $I^{MN}, X_{ij}^N$ defined previously. 

\section{Current amplitudes}

\subsection{$\langle J \mo \mo\rangle$ correlator}

To begin this section, we shall compute the three-point function of a current with two scalar operators using the embedding formalism. While the final result is well known, this calculation will serve to illustrate the usage of the embedding formalism for the computation of current amplitudes. Also, as we shall see in the next section, it will immediately give us the result for the three current amplitude.

We take for the gravitational action that of a minimally coupled scalar of mass $m^2=\Delta(\Delta-d)$,
	\be
	S=\int \ud^{d+1} x \sqrt{-g}\left(-\frac 14 F_{MN}F^{MN}+
	|\nabla_M \phi-i e A_M \phi|^2+m^2 \phi^2\right).
	\ee
The three point vertex is of the form
	\be
	i e\, A_M(P_3)(\nabla \phi(P_1) \phi(P_2)-\nabla \phi(P_2) \phi(P_1)).
	\ee
The amplitude is therefore
	\bea
	\langle J^M \mathcal O \mathcal O \rangle=2i e \, D_{d-1}^{M_3 A}
	\int \prod_{i=1}^3 \frac{\ud t_i}{t_i} t_i^{\Delta_i}
	\iads \!\!\ud X \left(t_1 P_{1,A}-t_2 P_{2,A}\right)
	\exp\left[2(t_1 P_1\!+\!t_2 P_2\!+\!t_3 P_3)\cdot X\right] \hspace{0.5 cm}
	\eea
with $\Delta_1=\Delta_2\equiv \Delta$. Recall that $D_{d-1}^{MA}$ is an operator which acts on the right-hand side of the expression. After the $X$ integration we obtain
	\be
	\langle J^M \mathcal O \mathcal O \rangle=
	2 i e \,\pi^h \, \Gamma\left(\frac{\sum_i \Delta_i+1-2h}2\right)
	D^{M_3A}_{d-1}
	\int \prod_{i=1}^3 \frac{\ud t_i}{t_i} t_i^{\Delta_i}
	\left(t_1 P_{1,A}-t_2 P_{2,A}\right)
	e^{-\sum_{i<j}^3t_i t_j P_{ij}}. \label{momcons}
	\ee
Let us focus on the integral. This is quite similar to the one we found for the scalar three point function, and we can proceed using a trick:
	\bea
	\int \prod_{i=1}^3 \frac{\ud t_i}{t_i} t_i^{\Delta_i}
	\left(t_1 P_{1,A}\right)
	e^{-\sum_{i<j}^3t_i t_j P_{ij}}=
	\int \prod_{i=1}^2 \frac{\ud t_i}{t_i} t_i^{\Delta_i}
	\frac{\ud t_3}{t_3} t_3^{d-2}
	\left(t_3t_1 P_{1,A}\right)
	e^{-\sum_{i<j}^3t_i t_j P_{ij}}&& \nonumber \\
	=-P_{1,A} \frac{\partial}{\partial P_{13}}\int \prod_{i=1}^2 \frac{\ud 	t_i}{t_i} t_i^{\Delta_i}	\frac{\ud t_3}{t_3} t_3^{d-2} e^{-\sum_{i<j}^3t_i t_j P_{ij}}=
	\frac{\delta_{13}}2 \frac{P_{1,A}}{P_{13}} \prod_{i<j} \Gamma(\delta_{ij})(P_{ij})^{-\delta_{ij}} &&.
	\eea
In the last expression, the $\delta_{ij}$ satisfy the constraint $\sum_{i\neq j}\delta_{ij}=\Delta_j-s_j$, with $s_j$ the spin of the field $j$.
Overall, we get
	\be
	\langle J^M \mathcal O \mathcal O \rangle=
	i e \pi^h \, \Gamma\left(\frac{\sum_i \Delta_i+1-2h}2\right)
	D^{M_3 A}_{d-1}
	\left(	
	\delta_{13} \frac{P_{1,A}}{P_{13}}-\delta_{23} \frac{P_{2,A}}{P_{23}}
	\right)
	\prod_{i<j} \Gamma(\delta_{ij})(P_{ij})^{-\delta_{ij}} 
	\ee
To finish, we are left with the action of the operator $D^{MA}_{d-1}$. However, this action is particularly simple here. To see this, first write
	\be
	D^{MA}_{d-1}=\eta^{M_3A}+\frac{1}{d-1}P_3^A\partial_{M_3}
	=\frac{d-2}{d-1}\, \eta^{M_3 A}+\frac{1}{d-1}
	\frac{\partial}{\partial P_3^{M_3}} P_3^A
	\ee
Since we have $\delta_{13}=\delta_{23}=\frac 12 (d-2)$, the second term in the operator leads to a vanishing result. Restoring the external leg normalizations the final answer is
	\bea
	\langle J^{M_3} \mathcal O \mathcal O \rangle
	&=&i e\, C
	\left(	
	\frac{P_{1}^{M_3}}{P_{13}}-\frac{P_{2}^{M_3}}{P_{23}}
	\right)
	\prod_{i<j}(P_{ij})^{-\delta_{ij}} 
	\eea
with
	\be
	C=\frac{1}{4 \pi^{2h}} \frac{\Gamma(h)\Gamma(\Delta)}{\Gamma(1+\Delta-h)^2}
	\ee
Following our general discussion in section \reef{confinv} we have $P_3^{M_3}\langle J^{M_3} \mathcal O \mathcal O\rangle=0$, and the index structure is indeed of the form $X_{12}^{M_3}$ as expected. 

\subsection{Current three-point amplitude}

We now consider a three point amplitude of a non-abelian Yang-Mills field in $AdS$, or alternatively, the conformal correlation function of three currents valued in some Lie algebra with structure constants $f_{abc}$. The Witten diagram is essentially same as in figure \ref{scalar3pt}. As usual, the $X$ integration is trivial and we can immediately write
	\bea
	&&\langle J^{a,M_1}(P_1)J^{b,M_2}(P_2)J^{c,M_3}(P_3)\rangle=i\,e\,\left(2\pi^h\right) \Gamma(d-1)f^{abc} \,D^{M_1 A}D^{M_2 B}D^{M_3 C} I_{ABC},\nonumber \\
	&&I_{ABC}=\int \prod_{i=1}^3 
	\frac{\ud t_i}{t_i} t_i^{\Delta_i}
	\left[
	\eta_{AB} \left(t_1 P_{1,C}-t_2 P_{2,C}\right)+\mbox{perms}
	\right]
	e^{-\sum_{i<j} t_i t_j P_{ij}}. \hspace{1 cm}
	\eea
This expression is remarkable, in that most of the complicated index structure has effectively been moved to the action of the $D$ operators. Each of the permutations inside the integral sign is essentially nothing but the three point function of a current with two scalars, which we have already computed! Therefore we can immediately write down
	\bea
	I^{ABC}=\frac{(d-2)^2}2\left(\frac{\X12^C \, \eta^{AB}}{P_{12}}+\mbox{perms}\right)
	\prod_{i<j} \Gamma(\delta_{ij}) (P_{ij})^{-\delta_{ij}}.
	\eea
where $\delta_{ij}=\frac 12 (d-2)$. To obtain the full amplitude one has but to mechanically act with the $D$ operators on the expression. Before we do this however, it is worth noticing the simplicity of the expression between parenthesis, which bears an uncanny resemblance to a gauge theory amplitude:
	\be
	\left(\frac{P_1^C}{P_{13}}-\frac{P_2^C}{P_{23}}\right) 				\frac{\eta^{AB}}{P_{12}} \quad \to \quad \left( k_1^c-k_2^c\right) 		\eta^{ab}
	\ee
Also, we haven't defined what the Mellin amplitude should be for the case of amplitudes involving currents. A natural definition seems to be that one should take as the Mellin amplitude the expression between parenthesis, since once this is given the entire real space amplitude may be determined after the action of the $D$ operators.

As a check that we haven't made a mistake, we may evaluate the full amplitude by acting with the $D$ operators. After some work one obtains:
	\bea
	&& \left\langle J^{a,M_1}(P_1)J^{b,M_2}(P_2)J^{c,M_3}(P_3)\right\rangle= \nonumber \\
	&& C_3 \left[\left(\frac{I^{M_1 M_2} \X12^{M_3}}{P_{12}}+\mbox{perms}\right)
	-\frac 32 \frac{d-2}{2d-3} \X12^{M_3}\X23^{M_1}\X31^{M_2}
	 \right]\prod_{i<j}^3 (P_{ij})^{-\delta_{ij}} \hspace{2cm}
	\eea
with 
\be
C_3=\frac{i e}{8 \pi^h} f^{abc}\Gamma(d-1)\frac{(2d-3)(d-2)}{(d-1)^3}.
\ee
This agrees with previous results in the literature \cite{D'Hoker:1999jc} up to normalization conventions. Also as expected, the full amplitude is a polynomial in $I,X$, and satisfies the ``gauge invariance'' condition. This calculation shows how the embedding formalism simplifies considerably the calculation of the amplitudes.

\subsection{Scalar 4-point with current exchange}

In this section we will be computing the contribution to the scalar 4-point function of a diagram where a gauge boson is being exchanged. This will be useful as practice to the calculation of the 4-current amplitude in the next section. It will also allow us to check our formalism is correct by checking that the pole structure of the Mellin amplitude agrees with the general results of Mack \cite{Mack:2009mi}.

The process we'll be considering is described by the Witten diagram in figure \ref{gaugeexchange}. 
\FIGURE[ht]{
	\parbox{10 cm}{
		\centering
		\includegraphics[width=7cm]{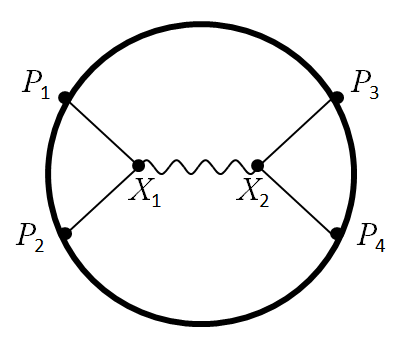}
		\caption{Gauge boson exchange diagram}.
	}
	\label{gaugeexchange}
}
The gauge-boson bulk-to-bulk propagator can be written as a product of two bulk-to-boundary propagators, and we can write
	\be
	A_4^J=\int \frac{\ud c}{2\pi i} f^1_{\delta}(c)\int \ud Q
	\langle J_{h+c}^M(Q) \mathcal O(P_1)\mathcal O(P_2)\rangle\eta_{MN}
	\langle J_{h-c}^N(Q) \mathcal O(P_3)\mathcal O(P_4)\rangle
	\ee
We have already computed the three-point functions appearing in the expression above. However, in practice one does not want work with the three-point function, but rather with its Schwinger parameterized form, as to be able to perform the $Q$ integral. 

Notice that in the three point functions above, the currents $J$ have conformal dimensions $h\pm c$, and not $d-1$ as usual; that is
	\bea
	&& \left \langle J_{h\pm c}^M(Q) \mathcal O(P_1)\mathcal O(P_2)\right 	\rangle=\nonumber
	 \\
	&&2i e\pi^h D^{MA}_{h\pm c} \int \frac{\ud t_1}{t_1} \frac{\ud t_2}{t_2}\frac{\ud s}s t_1^{\Delta_1} t_2^{\Delta_2} s^{h+c}(t_1 \!P_{1,A}-t_2\! P_{2,A})\exp\left[-t_1 t_2 P_{12}+2 s(t_1 P_1+t_2 P_2)\cdot Q\right] \hspace{1 cm}
	\eea
In order to perform the $Q$ integrals, we need to do something about the $Q$ and $Q$ derivative hidden in the $D$ operators. However, as in the calculation of $\langle J\mathcal O \mathcal O\rangle$ amplitude, we can write
	\be
	D_{h\pm c}^{MA}=\frac{h\pm c-1}{h\pm c} \eta^{MA}+ \frac{\partial}{\partial Q^M} Q^A,
	\ee
and, as before, the second term does not contribute. Each $D$ operator reduces to a Minkowski metric times a factor, and the contraction of both of them leads to
	\be
	\eta_{MN} D^{MA}D^{NB} \to \frac{(h-1)^2-c^2}{h^2-c^2}\eta^{AB}
	\ee
The prefactor in the above exactly cancels a similar factor in the definition of $f_{\delta,1}(c)$, reducing it to $f_{\delta,0}(c)$ ({\em c.f.} equation \reef{f1}). The $Q$ integrations proceed as in the scalar exchange computation of section \ref{scalarexchange}, and we get
	\bea
	&& A_4^J=e^2\,\left(8\pi^{3h}\right)
	\int_{-i \infty}^{+i\infty} \frac{\ud c}{2\pi i} 				f_{\delta,0}(c) \int \widetilde{\ud^2 s} \, \,
		\gt{1+2\Delta +c-h}\gt{1+2\Delta-c-h}\nonumber \\		
		&& \int \prod_{i=1}^4 \frac{\ud t_i}{t_i} t_i^{\dd i}
		J_1\cdot J_2 \exp\left[-(1+s^2)t_1 t_2 P_{12}-(1+\bar s^2)t_1 t_2 P_{34}
		-s \bar s\,\sum'_{(ij)} t_i t_j P_{ij}\right]. \label{j1j2}
	\eea
where we have defined the ``currents'':
 \be
  J_1=t_1 P_1-t_2 P_2, \qquad J_2=t_3 P_3-t_4 P_4.
 \ee
This expression is very close to the corresponding one for scalar exchange, and accordingly the rest of the calculation is now essentially the same.
Using Symanzik's star formula we write the above as a Mellin amplitude,

	\bea
		M(\delta_{ij})&=&
		8\gamma_{12} e^2 \, \int_{-i \infty}^{+i\infty} \frac{\ud c}{2\pi i} 	f_{\delta,0}(c)\,
		I(12,h-1,c) I(34,h-1,-c),
	\eea
with $\gamma_{12}=\frac{s_{13}-s_{23}}2$ and $I(12,h,c)$ as in \reef{I12}, except for the crucial difference $h\to h-1$. This difference arises from the extra factors of $1/s$, $1/\bar s$ in the integrals relative to the ones appearing in the Mellin amplitude for scalar exchange. These in turn appear due to the presence of the non-exponentiated $P_{13}, P_{24},\ldots$ terms in the integrand of \reef{j1j2}. After these integrals are performed we obtain
	\bea
		M(\delta_{ij})&=& 4\gamma_{12}\,e^2
		\, \int_{-i \infty}^{+i\infty} 
		\frac{\ud c}{2\pi i} \frac{l_{h-1}(c)l_{h-1}(-c)}{(\delta-h)^2-c^2}
	\eea
with $\delta=d-1$. To evaluate the integral we simply notice that it is the same as that appearing in a scalar exchange diagram of conformal dimension $\Delta=\delta-1=d-2$ and in dimension $h\to h-1$. Therefore we can evaluate it exactly to find
	\bea
	&& M(s_{12})=
	\frac{4\gamma_{12}}{s_{12}-(\delta-1)}\, \frac{e^2 \gt{2\Delta+\delta-h} \gt{2\Delta+\delta-h}}{\Gamma(1+\delta-h)} 
	\nonumber \\
	&& _3 F_2\left(\frac{1-2\Delta+\delta}2,
	\frac{1-2\Delta+\delta}2,\frac{(\delta-1)-s_{12}}2; \frac{1+\delta-s_{12}}2,1\!+\!\delta-\!h;1\right).
	\label{cur1}
	\eea	
Alternatively, we can find the poles in $s_{12}$ by pole pinching to find their position has shifted. The result is
	\be
M(s_{12})=\sum_{n=0}^{+\infty}\frac{4\gamma_{12} }{s_{12}-(\delta-1)-2n}\,	P_n^{\delta}\,\hat V_{[0,0,n]}^{\Delta,\Delta,\delta-1}\hat V_{[0,0,n]}^{\Delta,\Delta,\delta-1} \label{cur2}
	\ee
where it is understood that $\delta=d-1$, and we have
	\bea
	\hat V^{\Delta,\Delta,\delta-1}_{[0,0,0]}&=&
		e\, \gt{(\delta-1)+2\Delta-2(h-1)},  \\
	\hat V^{\Delta,\Delta,\delta-1}_{[0,0,n_1]}&=&
		\hat V^{\Delta,\Delta,\delta-1}_{[0,0,0]} 
		\left(1-\frac 12 \left[2\Delta+(\delta-1)\right]+(\delta-1)\right)_{n_1}  \\
	P_n^\delta&=& \left[n! \Gamma\left(1+\delta-h+n\right)\right]^{-1}.			
	\eea
There are several interesting things to notice in this result.
For instance these are essentially the same vertices appearing in $\phi^3$ theory, upon shifting $h\to h-1$, $\delta\to \delta-1$.
Also, this is an exact expression, i.e. there are no terms analytic in $s_{12}$ that we've missed, and expressions \reef{cur1}, \reef{cur2} are identical. The main novelty is the factor of $\gamma_{12}$, whose appearance however had already been predicted by Mack \cite{Mack:2009mi}. It is interesting to notice that the amplitude shows factorisation, since this term is given by
	\be
	2\gamma_{12}= s_{13}-s_{23}=(k_1-k_2)\cdot (k_3-k_4).
	\ee
More precisely, it would show exact factorisation if the $P$'s appearing in the index structures of the three-point amplitudes $\langle J\mo\mo\rangle$, could be transformed into $k$'s. The simplicity of this result suggests that our Feynman rules can be perhaps extended to the case where there are propagating currents. 

\subsection{Current 4-point amplitude}

In this section, we compute a four point function of currents using $AdS$/CFT. We consider non-abelian gague theory in $AdS$, described by an action
	\be
	S_{YM}=-\int \ud^{d+1}x\sqrt{-g}\, \frac 14\, \mbox{Tr}\left( F_{MN} F^{MN}\right)
	\ee
with $F_{MN}^a=\partial_M A_N^a-\partial_N A_M^a+i e f^{abc} A_M^b A_N^c$, and want to evaluate the CFT amplitude
	\be
	A_4=\left \langle J^{a,M_1}(P_1)J^{b,M_2}(P_2)J^{c,M_3}(P_3)J^{d,M_4}(P_4)\right 	\rangle
	\ee
From the action above, there are two kinds of diagrams contributing to the current four point function, a contact interaction and a current exchange diagram. The latter can occur in any of three different channels - we show the $s$-channel diagram in figure \ref{cur4pt}.
\FIGURE[ht]{
	\parbox{10 cm}{
	\centering
	\includegraphics[width=7cm]{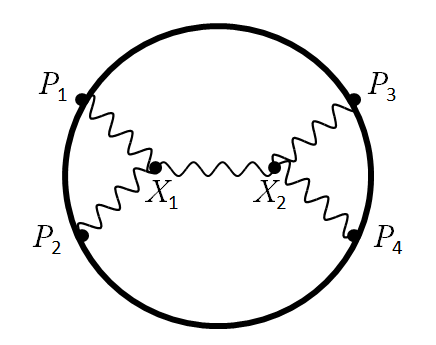}
	\caption{Current four-point function amplitude with current exchange.}
		}
	\label{cur4pt}
}
The contact interaction is elementary using our methods, since there is only an $X$ integration to perform which is trivial, and the amplitude is immediately written
	\bea
	A_c&=& \frac{\pi^h}2 \mathcal E_4 \int \ud \delta_{ij}\left(\prod_{i=1}^4 D_{d-1}^{M_i A_i}\right) 
	C_4\left[f^{abe}f^{cde}\eta_{A_1 A_3}\eta_{A_2 A_4}+\mbox{perms}\right] 
	\prod_{i<j}^4 \Gamma(\delta_{ij}) (P_{ij})^{-\delta_{ij}}\hspace{1 cm}
	\eea
where $\sum_{i\neq j} \delta_{ij}=d-1$ and the overall constant is
	$
	C_4=i e^2\gt{3d-4}.
	$
The $D$ operators act on the products of $P_{ij}$ and are contracted with the Minkowski metrics to give the overall index structure.  Notice that the integrand contains the Yang-Mills theory contact diagram in flat space. As a non-trivial check on the arguments of section \ref{confinv}, we show in appendix \ref{contact} that the result of acting with $D$ operators is indeed a polynomial in $I,X$ structures.
Let us move on to the exchange diagrams. In the following we shall only consider the $s$-channel exchange, and we will denote the corresponding amplitude by $A_s$. As usual, the four-point function is the gluing of two three-point functions,
	\bea
	A_s\!=\!	\int \frac{\ud c}{2\pi i} f^1_\delta\!\! \ib\!\! \!\!\ud Q 
	\left \langle J^{a,M_1}(P_1) J^{b,M_2}(P_2) J_{h+c}^{e,N}(Q) \right \rangle
	\left \langle J^{c,M_1}(P_1) J^{d,M_2}(P_2) J_{h-c}^{e,N}(Q) \right \rangle
	\eea
with
	\bea
	&&\left \langle J^{a,M_1}(P_1) J^{b,M_2}(P_2) J_{c,h+c}^N(Q)\right \rangle 	
	=e\,(2\pi^h)\,f^{abc}\int \frac{\ud t_1}{t_1}\frac{\ud t_2}{t_2}\frac{\ud s}{s}
	t_1^{d-1} t_2^{d-1} s^{h+c}\nonumber \\
	&&	
	\, D^{M_1 A_1} D^{M_2 A_2} D_{h+c}^{N A_3}
	\left[(t_1 P_1-t_2 P_2)_{A_3} \eta_{A_1 A_2}+
	(t_2 P_2-s Q)_{A_1}\eta_{A_2 A_3}
	\right. \nonumber \\
	&&+\left.(s Q-t_1 P_1)_{A_2}\eta_{A_1 A_3}\right]
	 \exp\left(-t_1 t_2 P_{12}+2 s (t_1 P_1+t_2 P_2)\cdot Q\right).
	 \label{temp}
	\eea
The presence of $Q$'s in the expression, and also of $Q$ derivatives inside the $D_{h+c}$ operator complicates the calculations. Fortunately, there is a significant simplification. Recall that originally we had $D^{MA} X_A=0$. After the $X$ integrations are performed this means that
	\be
	\int \left(\prod_i \frac{\ud t_i}{t_i} t_i^{\Delta_i}\right) \, D^{MA}\left(\sum t_i P_{i,A}\right) e^{-\sum t_i t_j P_{ij}}=0.
	\ee
We interpret this as ``momentum conservation''.
Now, the operator $D_{h+c}$ is given by
	\be
	D_{h+c}^{N A_3}=\frac{h+c-1}{h+c}\eta^{N A_3}+\frac{1}{h+c} \frac{\partial}{\partial Q^N} Q^{A_3}.
	\ee
Consider contracting the second piece of the above with each term of the second line of \reef{temp}. The first such term leads to a vanishing result, since it is nothing but the operator $D^{M_1 A_1}D^{M_2}_{\ A_1}$ acting on a $\langle J^{N} \mathcal O \mathcal O\rangle$ amplitude, which vanishes when contracted with $Q^N$. The remaining two terms on the second line become
	\be
	\simeq 
	t_2 P_{2,A_1} Q_{A_2}-t_1 P_{1,A_2} Q_{A_1}
	\ee
Using momentum conservation to trade $Q$ for $P_1$ and $P_2$ and the result is easily seen to vanish (recall that $D^{M_i A_i}P_{A_i}$ is vanishing). Therefore, in $D_{h+c}$ it suffices to keep its Minkowski metric part. Further, any $Q$ with a free index may be traded for $P_1, P_2$. The net result is that we have
\bea
&& \left \langle J^{a,M_1}(P_1) J^{b,M_2}(P_2) J_{h+c}^{c,N}(Q)\right \rangle 	=e\,(2\pi^h)\,f^{abc}\int \frac{\ud t_1}{t_1}\frac{\ud t_2}{t_2}\frac{\ud s}{s}
	t_1^{d-1} t_2^{d-1} s^{h+c}	
	\, D^{M_1 A_1} D^{M_2 A_2}\nonumber \\
&&
	\frac{h+c-1}{h+c}\, \eta^{N A_3} \left[(t_1 P_1-t_2 P_2)_{A_3} \eta_{A_1 A_2}+
	2 t_2 P_{2,A_1}\eta_{A_2 A_3}-2 t_1 P_{1,A_2}\eta_{A_1 A_3}\right] \nonumber \\
&&	\exp\left(-t_1 t_2 P_{12}+2s(t_1 P_1+t_2 P_2)\cdot Q\right).
	 \label{temp2}
	\eea
Of course, a completely analogous expression holds for the other three point function appearing in \reef{temp}. Since all the details of index structure have now decoupled from the integrals, the rest of calculation is essentially the same as that of the current exchange diagram of the previous section. The $Q$ integral is performed, and the result can be put into the form of a Mellin amplitude using Symanzik's star formula. In the end we obtain
	\be
	A_s=\frac{\pi^h}2\,\mathcal E_4\,\int \ud \delta_{ij} \left(\prod_{i=1}^4 D^{M_i A_i}\right) 			M_{A_1,\ldots, A_4}(\delta_{ij}) 
	\prod_{i<j} \Gamma(\delta_{ij}) (P_{ij})^{-\delta_{ij}}
	\ee
with
\bea
	&& M^{A_1 A_2 A_3 A_4}(s_{12})=\frac 12
	\frac{I^{A_1 A_2 A_3 A_4}(s_{12},\gamma_{12})  }{s_{12}-(d-2)}\, \frac{e^2 \gt{3(d-1)-h}^2}{\Gamma\left(\frac{d}2\right)} 
	\nonumber \\
	&& _3 F_2\left(\frac{2-d}2,
	\frac{2-d}2,\frac{d-2-s_{12}}2; \frac{d-s_{12}}2,\frac d2;1\right).
	\label{cur4}
	\eea
or equivalently,
	\be
	M^{A_1\ldots A_4}=I^{A_1 A_2 A_3 A_4}(s_{12},\gamma_{12})
	\sum_{n=0}^{+\infty}\frac{P_n^{d-1} }{s_{12}-(d-2)-2n} 		\, \hat V_{[0,0,n]}^{d-1,d-1,d-2}\hat V_{[0,0,n]}^{d-1,d-1,d-2}.
	\ee
The vertices in the above are the same that appeared in \reef{cur2}
Specializing our expressions for $d=4$ we get the simple result
	\be
	M^{A_1\ldots A_4}=\frac{225\pi e^2}{256} I^{A_1 A_2 A_3 A_4}(s_{12},\gamma_{12}) \left(\frac{2}{s-2}+\frac{1}{s-4}\right)
	\ee
We have yet to characterize the index structure $I_{A_1 A_2 A_3 A_4}$. It is the result of contracting two currents of the form
	\bea
	J_{A_i A_j A_k}\equiv 
	(t_i P_i-t_j P_j)_{A_k} \eta_{A_i A_j}+
	2 t_j P_{j,A_i}\eta_{A_j A_k}-2 t_i P_{i,A_j}\eta_{A_i A_k}.
	\eea
followed by $t_i t_j P_{ij}\to \frac{\delta_{ij}}{P_{ij}}$. Doing this we obtain
	\bea
	I^{A_1 A_2 A_3 A_4}&=&
	\,4 \gamma_{12}\, \eta^{A_1 A_2}\eta^{A_3 A_4} \nonumber \\
	&-&4 \left[ \frac{(\gamma_{12}-s_{12})}{2 P_{13}}\left(
	\eta^{A_3 A_4} P_1^{A_2} P_3^{A_1}+
	\eta^{A_1 A_2} P_1^{A_3} P_3^{A_4}
	-2 \eta^{A_1 A_3} P_1^{A_2}P_3^{A_4} \right)\right.\nonumber \\
	&& -(1\leftrightarrow 2)-(3\leftrightarrow 4)+(1\leftrightarrow 2,\, 3\leftrightarrow 4)\bigg].
	\eea
It is clear that if one identifies $P_i$ with a fictional momentum $k_i$, then the index structure of this expression roughly corresponds to the one appearing in the analogous diagram for Yang-Mills theory in flat space. To obtain the full conformal index structure we have to act with the $D$ operators. This is most simply performed with the aid of a computer\footnote{Notebooks are available upon request.}. The result is too long to be presented here, but we have been able to check that it is simply a polynomial in the $X_{ij}^M$ and $I^{MN}$ structures introduced in \reef{xdef}, \reef{idef}, as expected from our general arguments in section \ref{confinv}.

Importantly the propagator/vertex structure remains, and it is exactly the same as what we have computed in the scalar four point function current exchange diagram. In this sense, that computation already contains all the dynamic information relevant for the four-current correlator. What the current result shows is that it is possible to quite simply decouple the details of the index structure from the rest of the calculation.

\section{Discussion and Outlook}

In this paper we have showed how calculations of correlation functions in $AdS$/CFT are significantly made simpler by the combined use of the embedding formalism and the Mellin representation. The embedding formalism essentially makes the kinematic $AdS$ integrals become trivial, at the expense of introducing integrations in Schwinger parameters. At this point the Mellin representation becomes useful by translating such integrations to Mellin space via Symanzik's formula. With these methods we have managed to write down four point Mellin amplitudes explicitly in terms of hypergeometric functions. For higher point amplitudes, we have shown how there seems to be a set of Feynman rules which allows us to write them down. Although we have not proved in full generality that these rules are correct, we have presented non-trivial evidence in the form of the explicit calculation of higher point amplitudes.

The similarity between Mellin amplitudes and flat space scattering amplitudes had been noticed already in \cite{Penedones:2010ue}. There it was conjectured that in the high energy limit where the $\delta_{ij}$ parameters become large, the Mellin amplitude reduces to a flat-space amplitude of massless particles. In this sense, $AdS$ space can be thought of as naturally providing an IR cut-off for flat-space amplitudes. As far as we have been able to check, the results we have derived in this paper agree with the proposal of \cite{Penedones:2010ue}, at least in the scalar sector. When free indices are present, we are faced with difficulties, as the Mellin amplitude now depends on the coordinates $P$ as well as on the Mandelstam invariants. Our results suggest that we should identify the corresponding flat space amplitude with the reduced Mellin amplitude, i.e. the amplitude obtained before acting with the $D$ operators. Indeed, as we've pointed out throughout this paper, those amplitudes are remarkable similar to flat space amplitudes, if one identifies the coordinates $P$ with momenta $k$.

We clearly lack a deeper understanding of the structure of general Mellin amplitudes, such as pole structure, relation to lower point amplitudes and unitarity properties \footnote{For a proposed BCFW type recursion relation for Witten diagrams see \cite{Raju:2011ed,Raju:2011mp}.}
. Presumably such an understanding could lead to a proof of our proposed Feynman rules for Mellin amplitudes in scalar theory. It could also help us to understand the structure of amplitudes involving fields with spin, and if whether Feynman rules can be written down in this case. As a first easy check one should compute higher $n$-point functions of scalars with gauge fields propagating in the internal lines.

An obvious continuation of our work is the investigation of loop amplitudes. These were first discussed in \cite{Penedones:2010ue}, but there it was not attempted to write the result  {\em \`a la} Feynman. It would be interesting to check whether our rules for tree-level scalar amplitudes generalize to loop amplitudes in the expected way. Although in our formalism one would never obtain loop momenta integrals, one does obtain Mellin-Barnes type integrals, which roughly correspond to integrals over conformal dimension. Since the Mellin momenta $k_i$ square to conformal dimension, perhaps these integrals can be interpreted as integrals over the norm of the loop momenta.

Recently there was an attempt to use the spinor-helicity formalism to compactly describe CFT correlators in momentum space \cite{Maldacena:2011nz}. Our methods allow for a different tack on the same problem: since the embedding formalism allows us to describe the index structure of Mellin amplitudes in terms of $d+2$ vectors $P$ satisfying $P^2=0$, use of spinor-helicity formalism suggests itself. For instance one could to use the six-dimensional formalism of \cite{Cheung:2009dc} to describe four-dimensional conformal field theory amplitudes. Curiously, for $d=2$ it seems that the $\pm$ helicities of four dimensional massless particles map to (anti)holomorphic two-dimensional amplitudes. This is possible because after the action of $D$ operators, the conformal index structure of a CFT amplitude resembles that of a flat-space amplitude with higher dimension operators: the current 3-pt function has contributions cubic in $P$, which would come from an $(F_{ab})^3$ term in four dimensions.

It seems likely that the calculation of the stress-tensor four-point function should be achievable using our methods. The results we have obtained in this paper for the current four-point function lead us to expect that the index structure should decouple from the exchange part of the amplitude. The latter should essentially be the same as that obtained as for stress-tensor exchange in scalar theory. The full amplitude will be obtained by acting with four $D_2$ operators on the reduced Mellin amplitude, which should have an index structure similar to a four-graviton flat-space amplitude upon identification of the momentum with the coordinate $P$. We hope to present more on this and other stress-tensor correlation functions elsewhere \cite{progress}.

Finally, we have seen that there seems to be an intriguing connection between the correlation functions we have been computing for $d$-dimensional CFT's, and a theory of massless particles in $d+2$ dimensions. The connection is given by interpreting boundary point of the CFT as $d+2$ null vectors $P$, which can then be interpreted as momenta. It is highly suggestive that we were able to write down the relations \reef{gaugecond} and even a ``momentum conservation'' equation \reef{momcons}. It would be interesting to see if this connection can be developed further.

\acknowledgments
It is a pleasure to acknowledge discussions with Atish Dabholkar, Paolo Benincasa, Eduardo Conde and Xi\`an Camanho. The author would like to thank the University of Santiago de Compostela, where part of this work was performed, for funding and hospitality. 
The author acknowledges funding from the LPTHE, Universit\'e Pierre et Marie Curie, and partial support from the Portuguese FCT funded project CERN/FP/116377/2010.

\appendix
\section{Some integrals}

In this section we describe the computation of the $AdS$ and $AdS$ boundary integrals which appear throughout the paper. These calculations have appeared already in \cite{Penedones:2010ue}, and we include them here for completeness. The first such calculation is the proof that
	\be
	\int_{0}^{+\infty} \prod_{i}\left(\frac{\ud t_i}{t_i} t^{\alpha_i}\right) \iads \ud X\, e^{2 T\cdot X}
	=
	\pi^h \gt{\sum_i \alpha_i-2h} 
	\int_{0}^{+\infty} \prod_{i}\left(\frac{\ud t_i}{t_i}  t^{\alpha_i}\right) e^{T^2}. \label{Xint}
	\ee
with $T=\sum t_i P_i$. We proceed by computing the left-hand side. First we evaluate the $AdS$ integral. By Lorentz invariance we can consider the case where $T=|T|(1,1,0)$. We also parameterize $AdS_{d+1}$ space by
	\be
	X=(X^+,X^-,X^\mu)=\frac{1}{x_0}\, (1,x_0^2+x^2,x^\mu)
	\ee
and define $h\equiv d/2$.
Then we get
	\bea
	\iads \ud X\, e^{2 T\cdot X}&=& \int_0^{+\infty} 
	\frac{\ud x_0}{x_0}\, x_0^{-d} \int_{0}^{+\infty} \ud^d x 
	\, e^{-(1+x_0^2+x^2)|T|/x_0} \nonumber \\
	&=& \pi^h \int_0^{+\infty} 
	\frac{\ud x_0}{x_0}\, x_0^{-h}\, e^{-x_0+T^2/x_0}
	\eea
The original integral becomes
	\bea
	&&\pi^h \int_{0}^{+\infty} \prod_{i}\left(\frac{\ud t_i}{t_i} t^{\alpha_i}\right) \int_0^{+\infty} 
	\frac{\ud x_0}{x_0}\, x_0^{-h}\, e^{-x_0+(\sum_i t_i P_i)^2/x_0}
	= \nonumber \\
	&=& \pi^h \int_{0}^{+\infty} \prod_{i}\left(\frac{\ud t_i}{t_i}  t^{\alpha_i}\right) e^{T^2}\int_0^{+\infty}\frac{\ud x_0}{x_0} \, x_0^{\sum_i \alpha_i/2-h}e^{-x_0}= \nonumber \\
	&=& \pi^h \gt{\sum_i \alpha_i-2h} 
	\int_{0}^{+\infty} \prod_{i}\left(\frac{\ud t_i}{t_i}  t^{\alpha_i}\right) e^{T^2}.
	\eea
where in the second step we rescaled $t_i\to t_i/\sqrt{x_0}$

Next we prove:
	\bea
	\int_{0}^{+\infty} \frac{\ud s}{s} \frac{\ud \bar s}{\bar s} 	s^{h+c}s^{h-c}\,
	\ib \ud Q\, e^{2 T\cdot Q}
	=2\pi^h \int_{0}^{+\infty} \frac{\ud s}{s} \frac{\ud \bar s}{\bar s} s^{h+c}s^{h-c} e^{T^2}	\label{boundint}
	\eea
with $T\equiv (s X+\bar s Y)$.
First we evaluate the boundary integral on the left-hand side. Using the parameterization
	\be
	Q=(Q^+,Q^-,Q^\mu)=(1,x^2,x^\mu)
	\ee
we find
	\be
	\ib \ud Q\, e^{2 T \cdot Q}=\int_{0}^{+\infty} \ud^d x
	\, e^{-|T|(1+x^2)}=\frac{\pi^h}{|T|^h}\, e^{-|T|}.
	\ee
Now, noticing that $1=\int_{0}^{+\infty} \ud v\, \delta(v-s-\bar s)$, we find
	\bea
	&& \int_{0}^{+\infty} \frac{\ud s}{s} \frac{\ud \bar s}{\bar s} 	s^{h+c}s^{h-c}\, \frac{\pi^h}{|T|^h}\, e^{-|T|}= \nonumber \\
	&=& \int_{0}^{+\infty} \ud v
	\int_{0}^{+\infty} \frac{\ud s}{s} \frac{\ud \bar s}{\bar s} 	s^{h+c}s^{h-c}\delta(v-s-\bar s) \, \frac{\pi^h}{|s X+\bar s Y|^h}\, e^{-|s X+\bar s Y|} \nonumber \\
	&=& \pi^h\int_{0}^{+\infty} \frac{\ud v}v
	\int_{0}^{+\infty} \frac{\ud s}{s} \frac{\ud \bar s}{\bar s} 	s^{h+c}s^{h-c}\delta(1-s-\bar s) \, \frac{v^h}{|s X+\bar s Y|^h}\, e^{-|s X+\bar s Y|} \nonumber \\
	&=& \pi^h\int_{0}^{+\infty} \frac{\ud v}v
	\int_{0}^{+\infty} \frac{\ud s}{s} \frac{\ud \bar s}{\bar s} 	s^{h+c}s^{h-c}\delta(1-s-\bar s) \, v^h \, e^{v(s X+\bar s Y)^2} 
	\eea
Finally rescaling $s\to s/\sqrt v, \bar s \to \bar s/\sqrt v$ the $v$ integral is performed and we find the right-hand side of \reef{boundint}, as promised.
\section{The Symanzik star formula}
\label{Symanzik}

For completeness, in this section we review the Symanzik star integration formula in Euclidean space as discussed in \cite{Mack:2009mi}. For a proof and more details we refer the reader to the original reference \cite{Symanzik:1972wj}. Consider a set of $n$ points in Euclidean space $x_i$ and their differences $x_i-x_j$. In the embedding formalism we have $P_{ij}\equiv -2 P_i\cdot P_j=(x_i-x_j)^2$. Then Symanzik's formula is:
	\bea
	 \int_{0}^{+\infty}\!\! \left(\prod_{i=1}^n \frac{\ud t_i}{t_i} t^{\Delta_i}\right) e^{-\left(\!\!\sum_{1\leq i<j \leq n}\!\! t_i t_j\, P_{ij}\right)}&=&
		\frac{\pi^h/2}{(2\pi i)^{\frac 12 n(n-3)}}\int \ud \delta_{ij}\! \prod_{1\leq i<j \leq n}\!\!\Gamma(\delta_{ij})\,(P_{ij})^{-\delta_{ij}}
	\eea
The integration measure on the right-hand side deserves further explanation. The parameters $\delta_{ij}$, symmetric in $i,j$, satisfy
\be
\sum_{i\neq j} \delta_{ij}=\Delta_j \label{dij}
\ee
for all $i$. Now pick a particular solution of the set of equations \reef{dij}, $\delta^0_{ij}$.  Then we write
	\be
	\delta_{ij}=\delta^0_{ij}+\sum_{k=1}^{\frac 12 n(n-3)}\,c_{ij,k} s_k
	\ee
with
	\be
	c_{ii,k}=0,\qquad \sum_{j\neq i} c_{ij,k}=0.
	\ee
Choosing as independent coefficients the $\left (\frac 12 n(n-3)\right)^2$ coefficient $c_{ij,k}$ with $2\leq i < j \leq n$ (with the exception of $c_{23,k}$), with the further restriction $|\mbox{det} \, c_{ij,k}|=1$, we can write
	\be
	\int \ud \delta_{ij}\to \int_{-i \infty} \prod_{k=1}^{\frac 12 n(n-3)} \frac{\ud s_k}{2\pi i}
	\ee
The integration paths are chosen parallel to the imaginary axis, with real parts such that the real parts of the arguments of the gamma functions are positive.

\section{Details on the calculation of the six-point amplitude}
\label{details}

In the computation of the six point amplitude, or indeed of any amplitude involving internal lines, we have to perform boundary integrals over the coordinates $Q_i$ of each internal line. Since these integrals have to be done in a certain order, this breaks the symmetry of the expressions and the result seems more complicated than it is. A typical example of this is what happens in going from the third line of \reef{intfive} to the simpler looking \reef{5ptnice}. In the calculation of the six- and twelve-point functions the same thing occurs. In this section we give some details on the changes of variables required to obtain a simpler looking integral for the case of the six-point amplitude. Details on the 12-point amplitude are quite technical and can be obtained upon request.

We have six integrals that can be performed, over parameters $s_i,\bar s_i$, $i=1,\ldots 3$. After the $s_i$ integrations are performed, then if the boundary integrations were done in the order $Q_1, Q_2, Q_3$ the integral over the $\bar s_i$ is of the form
\bea
&&\int_0^{+\infty} \prod_{i=1}^3\left( \frac{\ud \bs i}{\bs i}
	\right)
\left(\bst 1+1\right)^{\frac{1}{2}
   (s_{12}-s_{34}-s_{56})}
   \left(\left(\bst 1+1\right) \bst 2+1\right)^{\frac{1}{2}
   (-s_{12}+s_{34}-s_{56})} \nonumber \\
  && \times \left(\bst 1
   \left(\bst 3 \left(\left(\bst 1+1\right)
   \bst 2+1\right)^2+\bst 2\right)+1\right)^{\frac{1}{2}
   (-c_1-h+s_{12})} \left(\left(\bst 1+1\right)^2
   \bst 2 \bst 3+1\right)^{\frac{1}{2}
   (-c_2-h+s_{34})}\nonumber \\
   && \times \left(\left(\bst 1+1\right)
   \bst 3 \left(\left(\bst 1+1\right)
   \bst 2+1\right)+1\right)^{\frac{1}{2}
   (-s_{12}-s_{34}+s_{56})},
   \eea
which looks quite complicated. However, performing the change of variables
	\be
	\bs 1\to \sqrt{x}, \quad \bs 2\to \sqrt{y}, \quad \bs 3\to \sqrt{z}
	\ee
followed by the sequence of variable changes
	\bea
	y\to \frac{y}{1+x}, \quad z\to \frac{z}{1+x}, \nonumber \\
	x\to \frac{x}{(1+y)(1+z)}, \quad y\to \frac{y}{1+z},
	\eea
finally leads to 
\bea
	\int_{0}^{+\infty}\frac{\ud x}{x}\frac{\ud y}{y}\frac{\ud z}{z}
x^a y^b z^c (1+x)^d(1+y)^e(1+z)^f(1+x+y+z)^g \ \hspace{1.5 cm}, && 
\eea
with $g=\frac 12 (c_1+c_2+c_3-h)$ and
	\bea
	a=\frac 12 (-c_1+h-s_{12}), \quad 
	b=\frac 12 (-c_2+h-s_{34}) \quad
	c=\frac 12 ( -c_3+h-s_{56}) &&\nonumber \\
	d=\frac 12 (-c_1-h+s_{12}), \quad
	e=\frac 12 (-c_2-h+s_{34}), \quad
	f=\frac 12 (-c_3-h+s_{56}). &&
	\eea
%

%

\section{Index structure of current four-point function contact diagram}
\label{contact}

We wish to evaluate:
	\be
	D^{M_1 A_1} D^{M_2 A_2}D^{M_3 A_3}D^{M_4 A_4}\left(
	\eta_{A_1 A_3}\eta_{A_2 A_4} \prod_{i<j} \left(P_{ij}\right)^{-\delta_{ij}}\right).
	\ee
Defining the quantities
	\begin{subequations}
	\begin{align}
	Z^{M_1}&\equiv s_{12}\left(X_{23}^{M_1}+X_{24}^{M_1}\right)
	+ \gamma_{12} X_{34}^{M_1}  \\
	Z^{M_2}&\equiv  s_{12}\left(X_{13}^{M_2}+X_{14}^{M_2}\right)
	-\gamma_{12} X_{34}^{M_2}  \\
	Z^{M_3}&\equiv  s_{12}\left(X_{14}^{M_3}+X_{24}^{M_3}\right)
	-\gamma_{12} X_{12}^{M_3}  \\
	Z^{M_4}&\equiv s_{12}\left(X_{13}^{M_4}+X_{23}^{M_4}\right)
	+\gamma_{12} X_{12}^{M_4}  \\
	\hat I^{M_i M_j}&\equiv 16 \delta_{ij} \frac{I^{M_i M_j}}{P_i\cdot P_j},
	\end{align}
	\end{subequations}
the result is
\bea
&& 256 (d-1)^4\left(P_{ij}\right)^{\delta_{ij}}	\left( D^{M_1 A_1} D^{M_2 A_2}D^{M_3 A_3}D^{M_4 A_4}
	\eta_{A_1 A_3}\eta_{A_2 A_4} \prod_{i<j} \left(P_{ij}\right)^{-\delta_{ij}}\right)= \nonumber \\
&& Z^{M_1}Z^{M_2}Z^{M_3}Z^{M_4}
+\left[\frac{(d-1)^2-\delta_{12}}{\delta_{12}}\right]^2
\hat I^{M_1 M_2}\hat I^{M_3,M_4} %
+\hat I^{M_1 M_3}\hat I^{M_2 M_4}
+\hat I^{M_1 M_4}\hat I^{M_2 M_3}\nonumber \\
&& +\left[\frac{(d-1)^2-\delta_{12}}{\delta_{12}}\right]
\left(Z^{M_3 M_4} \hat I^{M_1 M_2}+Z^{M_1 M_2} \hat I^{M_3 M_4}\right) \nonumber \\
&&+\left(Z^{M_1 M_3} \hat I^{M_2 M_4}+
Z^{M_1 M_4} \hat I^{M_2 M_3}+
Z^{M_2 M_3} \hat I^{M_1 M_4}+
Z^{M_2 M_4} \hat I^{M_1 M_3}
\right),
\eea
in exact agreement with the expectations of section \ref{confinv}.

\bibliography{Biblio}{}
\bibliographystyle{JHEP}
\end{document}